\documentclass[sigconf]{acmart}

\usepackage{balance}
\usepackage{booktabs} 
\usepackage{pdfpages}
\usepackage{algorithm}
\usepackage{algorithmic}
\usepackage[color=green]{todonotes}
\usepackage{multirow}
\usepackage{caption}
\usepackage[utf8]{inputenc}
\usepackage{booktabs}

\usepackage{pgfplots}
\usepackage{graphicx}
\usepackage{caption}
\usepackage{subcaption}
\usepackage{pgfplots, pgfplotstable}
\pgfplotsset{compat=1.14}
\usetikzlibrary{pgfplots.groupplots,shapes,decorations}

\newlength\figureheight
\newlength\figurewidth
\newlength\smallfigureheight
\newlength\smallfigurewidth
\newlength\largefigureheight
\newlength\largefigurewidth
\newcommand\MarkerSize{2.0}
\newcommand\LineWidth{1.5}

\newenvironment{smitemize}{
\begin{itemize}
  \setlength{\topsep}{-3pt}
  \setlength{\itemsep}{1pt}
  \setlength{\parskip}{0pt}
  \setlength{\parsep}{0pt}
  \setlength{\leftmargin}{1em}  \setlength{\itemindent}{-4pt}
}{\end{itemize}}

\newcommand{\paragraphX}[1]{\vskip 4pt \noindent \textit{#1} \hskip .05in}
\begin{abstract}
Website fingerprinting enables a local eavesdropper to determine which websites a user is visiting over an encrypted connection. State-of-the-art website fingerprinting attacks have been shown to be effective even against Tor. Recently, lightweight website fingerprinting defenses for Tor have been proposed that substantially degrade existing attacks: WTF-PAD and Walkie-Talkie. In this work, we present Deep Fingerprinting (DF), a new website fingerprinting attack against Tor that leverages a type of deep learning called Convolutional Neural Networks (CNN) with a sophisticated architecture design, and we evaluate this attack against WTF-PAD and Walkie-Talkie. The DF attack attains over 98\% accuracy on Tor traffic without defenses, better than all prior attacks, and it is also the only attack that is effective against WTF-PAD with over 90\% accuracy. Walkie-Talkie remains effective, holding the attack to just 49.7\% accuracy. In the more realistic open-world setting, our attack remains effective, with 0.99 precision and 0.94 recall on undefended traffic. Against traffic defended with WTF-PAD in this setting, the attack still can get 0.96 precision and 0.68 recall. These findings highlight the need for effective defenses that protect against this new attack and that could be deployed in Tor.
.
\end{abstract}

\begin{document}

\title{Deep Fingerprinting: Undermining Website Fingerprinting Defenses with Deep Learning} 

\author{Payap Sirinam}
\orcid{}
\affiliation{%
  \institution{Rochester Institute of Technology}  
  \city{Rochester}
  \state{New York}  
}
\email{payap.sirinam@mail.rit.edu}

\author{Mohsen Imani}
\orcid{}
\affiliation{%
  \institution{University of Texas at Arlington}  
  \city{Arlington}
  \state{Texas}  
}
\email{mohsen.imani@mavs.uta.edu}

\author{Marc Juarez}
\orcid{}
\affiliation{%
  \institution{imec-COSIC KU Leuven}  
  \city{Leuven}
  \country{Belgium}  
}
\email{marc.juarez@kuleuven.be}

\author{Matthew Wright}
\orcid{}
\affiliation{%
  \institution{Rochester Institute of Technology}  
  \city{Rochester}
  \state{New York}  
}
\email{matthew.wright@rit.edu}

\settopmatter{printacmref=false, printccs=true, printfolios=true} 

\begin{abstract}
Website fingerprinting enables a local eavesdropper to determine which websites a user is visiting over an encrypted connection. State-of-the-art website fingerprinting attacks have been shown to be effective even against Tor. Recently, lightweight website fingerprinting defenses for Tor have been proposed that substantially degrade existing attacks: WTF-PAD and Walkie-Talkie. In this work, we present Deep Fingerprinting (DF), a new website fingerprinting attack against Tor that leverages a type of deep learning called Convolutional Neural Networks (CNN) with a sophisticated architecture design, and we evaluate this attack against WTF-PAD and Walkie-Talkie. The DF attack attains over 98\% accuracy on Tor traffic without defenses, better than all prior attacks, and it is also the only attack that is effective against WTF-PAD with over 90\% accuracy. Walkie-Talkie remains effective, holding the attack to just 49.7\% accuracy. In the more realistic open-world setting, our attack remains effective, with 0.99 precision and 0.94 recall on undefended traffic. Against traffic defended with WTF-PAD in this setting, the attack still can get 0.96 precision and 0.68 recall. These findings highlight the need for effective defenses that protect against this new attack and that could be deployed in Tor.
.
\end{abstract}


\maketitle

\section{Introduction}
With more than two million daily users, Tor has emerged as the de facto tool to anonymously browse the Internet~\cite{tor-metrics}. 
Tor is, however, known to be vulnerable to traffic analysis. In particular, {\em website fingerprinting (WF)} is a traffic analysis attack with the potential ability to break the privacy that Tor aims to provide. WF allows the attacker to identify web pages in an encrypted connection by analyzing patterns in network traffic. This allows a local and passive network adversary, such as a user's Internet service provider or someone sniffing the user's wireless connection, to identify the websites that the user has visited despite her use of Tor. 

WF exploits the fact that differences in website content (e.g., different images, scripts, styles) can be inferred from network traffic, even if traffic has been encrypted. From a machine learning perspective, WF is a classification problem: the adversary trains a classifier on a set of sites, extracting network traffic features that are unique to each website. To deploy the attack, the adversary uses the classifier to match traces of a victim to one of those sites. The effectiveness of WF depends heavily on both the classifier algorithm and the set of features used. Previous WF attacks use a set of hand-crafted features to represent Tor traffic, achieving 90\%+ accuracy against Tor using classifiers such as Support Vector Machine (SVM)~\cite{Panchenko2016}, $k$-Nearest Neighbors ($k$-NN)~\cite{Wang2014}, and random forests~\cite{Hayes2016}.


In response to these attacks, a number of defenses have been proposed. WF defenses add dummy packets into the traffic and add delays to real packets, aiming to hide features exploited by WF attacks such as traffic bursts and packet lengths. Notably, Tor Project developers have shown an interest in deploying {\em adaptive padding} as a possible defense~\cite{Perry2013, Perry2015}. Based on this, Juarez et al.\ proposed WTF-PAD and showed that it effectively defends against WF attacks with reasonable overheads, such that it would be practical for deployment in Tor~\cite{JR2016}. Recently, Wang and Goldberg proposed another effective and low-overhead defense called Walkie-Talkie (W-T)~\cite{Wang2017}. These proposals raise the question of whether attacks could be improved to undermine the effectiveness of the new defenses, a question we address in this work.


While the state-of-the-art attacks use classifiers that are popular in many applications, {\em deep learning} (DL) has shown to outperform traditional  machine learning techniques in many domains, such as speech recognition, visual object recognition, and object detection~\cite{LeCun2015}. Furthermore, DL does not require selecting and fine-tuning features by hand~\cite{Rimmer2018}. In this work, we thus explore whether we can leverage deep learning to improve classification results against defended Tor traffic. The key contributions of our work are as follows:

\begin{smitemize}
    \item We propose Deep Fingerprinting (DF), a new WF attack based on a Convolutional Neural Network (CNN) designed using cutting-edge DL methods. The attack uses a simple input format and does not require handcrafting features for classification. We describe how DF leverages advances from computer vision research for  effective and robust classification performance.

    \item To study the attack in detail, we experiment in a closed-world setting using a new dataset that we collected with 95 sites and 1,000 traces per site. We find that our DF WF attack is more accurate against Tor than the state-of-the-art attacks with 98.3\% accuracy. We also show results for how the number of training epochs and training dataset size affect the classification accuracy.

    \item We then show the effectiveness of the DF attack in the closed-world setting against Tor traffic defended with WTF-PAD and W-T. Against WTF-PAD, the attack reaches 90\% accuracy, which is significantly better than all other attacks. Against W-T, the attack reaches 49.7\% accuracy, which is better than all other attacks and nearly the theoretical maximum accuracy~\cite{Wang2017}.

    \item To investigate in a more realistic setting, we use an open world with 20,000 unmonitored sites. On non-defended traffic, the attack achieves 0.99 precision and 0.94 recall. On traffic defended with WTF-PAD, the attack yields 0.95 precision and 0.70 recall. We also examine the possibilities for attacking weak implementations of W-T.

    \item Based on our experimental findings, we propose a number of new directions to explore in both attack and defense.
\end{smitemize}

Overall, we find that the new DF WF attack undermines at least one defense that had been considered seriously for deployment in Tor~\cite{Perry2013,Perry2015}. We have disclosed our findings to the Tor Project, and they have expressed their concerns about WTF-PAD, setting the stage for more exploration of the design of realistic defenses.

\section{Threat Model}\label{threatmodel}

Among its goals, Tor aims to protect users against local eavesdroppers from learning what sites the user is going to. WF attacks, however, use traffic analysis to undermine Tor's protections. Prior work has shown that, under certain conditions, a local and passive adversary can identify the pages visited by a Tor user by exploiting patterns in network traffic~\cite{Herrmann2009,Panchenko2011,Cai2012,Wang2013,Wang2016,Panchenko2016,Hayes2016}.

To deploy the attack, the adversary captures the sequence of packets, also known as a \emph{traffic trace}, from each of a series of his own visits to a representative set of websites, including sites he is interested in detecting. From each trace, he then extracts features that are unique to each website. In the WF literature, we find a myriad of such features: packet size frequencies~\cite{Herrmann2009}, total transmission time and volume in both directions~\cite{Panchenko2011}, edit-distance score~\cite{Cai2012,Wang2013}, and the number of traffic bursts in each direction~\cite{Panchenko2011,Wang2014}, just to mention a few. As a result, the adversary obtains several \emph{feature vectors} for each website that are used to train a supervised classifier that learns how to identify the site from its features. Finally, the adversary can collect new traffic traces from the user's connection to the Tor network, extract the features, and use the trained classifier to guess the website.

In this work, we assume a network-level adversary that is: \textit{local}, meaning that he has access only to the link between the user and the entry node to the Tor network, and \textit{passive}, i.e., he can record network packets but not modify, delay, drop or decrypt them. Potential adversaries that might be in a position to deploy a WF attack include: eavesdroppers on the user's local network, local system administrators, Internet Service Providers (ISP), Autonomous Systems (AS) between the user and the entry node, and the operators of the entry node.

\begin{figure}[t]
  \centering
  \includegraphics[width=0.37\textwidth]{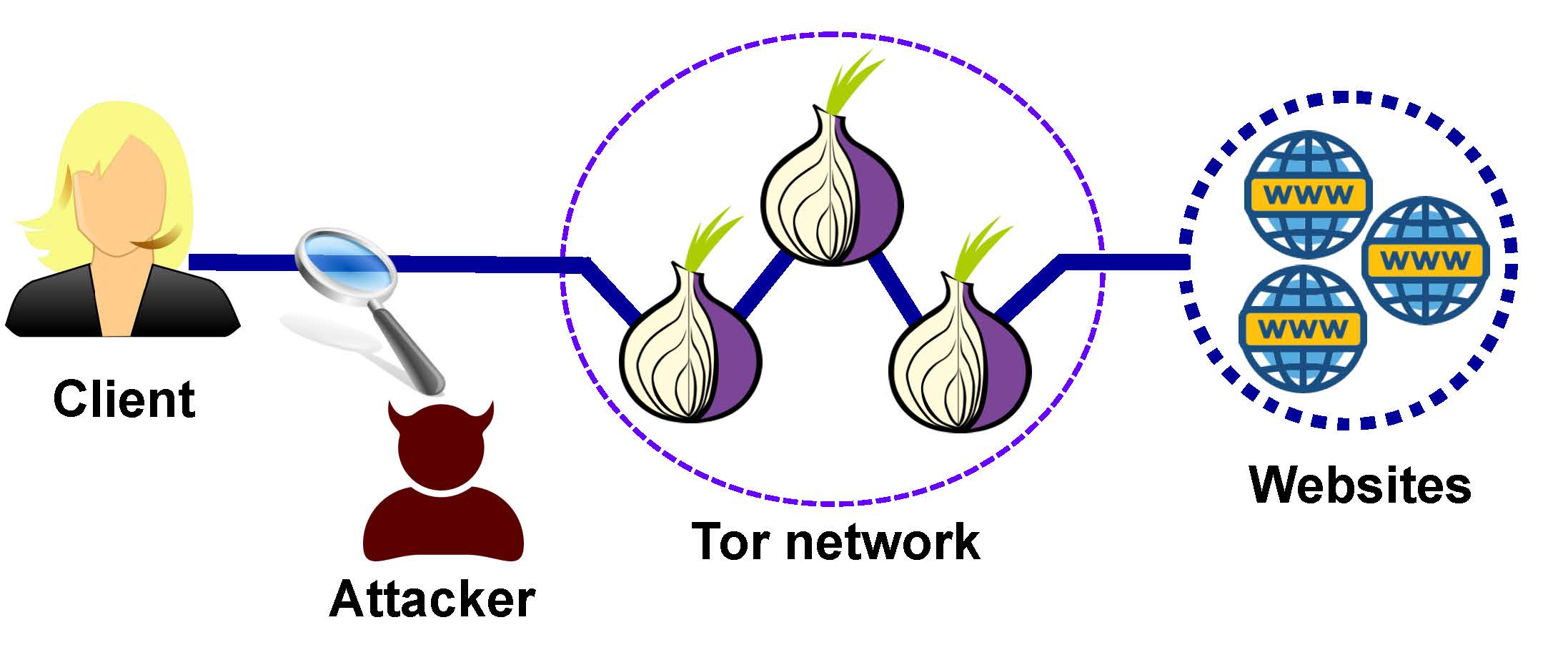}
  \caption{The WF threat model}\label{fig:threat_model}
\end{figure}

Figure~\ref{fig:threat_model} depicts the attack scenario: the client surfs the Web over the Tor anonymity system and the attacker intercepts the traffic between the client and the Tor network. We assume the adversary knows the client's identity and only aims at identifying the website. Note that the adversary can trivially obtain the client's IP address as long as he has access to the TLS connection between the user and the entry node. Beyond the entry node, Tor has stripped a layer of encryption and the IP of the client is no longer present in the headers of network packets.

Within this scenario, we draw on prior work to make several assumptions about the attacker goals and capabilities.

\paragraphX{Closed- vs Open-world Scenario:}
A closed-word assumes the user can only visit a small set of sites and that the adversary has samples to train on all of them~\cite{Cheng1998,Sun2002,Hintz2003,Herrmann2009}. This assumption was criticized for being unrealistic~\cite{Perry2013,Juarez2014}, as the world of sites that can be potentially visited is so large that not even the most powerful adversaries have the resources to collect data and train for every site. Subsequent studies have considered an open-world scenario, a more realistic setting in which the adversary can only train on a small fraction of the sites the user can visit. We use closed-world experiments for detailed comparison of different algorithms and parameter settings, and we report the results of open-world experiments for a more realistic evaluation of the attack.
In the open world, we follow the terminology used in prior work: the \textit{monitored set} includes sites that the adversary is interested in detecting, while the \textit{unmonitored set} are all other sites.

\paragraphX{Website vs Webpage Fingerprinting:}
In an abuse of language, most authors in the field use ``website fingerprinting'' to refer to the fingerprinting of only the home page of those websites. There is research that has attempted to fingerprint pages that are linked from the homepage~\cite{Cai2012}, but virtually all studies on website fingerprinting train and test the attacks on home pages. For comparison with prior work we make same assumptions in our evaluation.

\paragraphX{Traffic Parsing:}
As pointed out by Juarez et al.~\cite{Juarez2014}, the attacker is assumed to be able to parse all the traffic generated by a web visit and isolate it from other traffic (e.g., traffic generated by visits in other tabs, non-HTTP traffic over Tor, and so on). We note that the adversary is able to do so only if he deploys the attack from an entry node under his control. In that case, the adversary can select a domain's traffic by its Tor circuit ID. Concurrent and subsequent visits to the same domain would still go through the same circuit. If the adversary is eavesdropping the link between the client and the entry, all Tor traffic is multiplexed in the TLS connection to the entry. However, recent research has developed techniques to parse visits from multiplexed TLS traffic~\cite{Wang2016}. As with prior work, we assume that such parsing has already been done or is not needed.

\setlength{\parskip}{0pt}
 
\section{Background and Related Work}\label{sec:related}
In this section, we categorize and summarize prior work on WF attacks and defenses and then give the necessary background on deep learning to follow the rest of the paper.

\subsection{WF Attacks}


Herrmann et al.\ were the first to evaluate WF against Tor~\cite{Herrmann2009}. However, they only achieved 3\% accuracy in a closed world of 775 sites. The main problem with their approach was their reliance on packet length frequencies -- Tor sends data in fixed-size (512-byte) packets known as cells -- which renders this feature useless for classification of Tor traffic. In 2011, Panchenko et al.\ devised new features and improved the attack to 55\% accuracy on Herrmann et al.'s dataset~\cite{Panchenko2011}. Since then, the success rate of WF attacks against Tor has been incrementally improved, reaching 90\% accuracy by two classifiers using edit-distances~\cite{Cai2012,Wang2013}. These attacks, however, imposed high computational costs on the adversary, which makes them impractical for real-world deployment.


Recently, a new series of WF attacks have been proposed with advanced feature sets and more sophisticated classifiers that maintain the accuracy at 90\% while reducing the cost of the attack~\cite{Wang2014,Panchenko2016,Hayes2016}. These attacks have become the state-of-the-art WF attacks and are used to benchmark other attacks and defenses. We have selected them in this study to compare against our deep-learning-based DF attack. 

\paragraphX{$k$-NN.}
Wang et al.~\cite{Wang2014} proposed the $k$-NN attack. This approach consists in applying a $k$-Nearest Neighbors ($k$-NN) classifier, including features such as packet ordering, number of incoming and outgoing cells and numbers of bursts. These features are used in combination to form a distance metric (e.g., Euclidean distance) to measure the similarity between different websites. $k$-NN exhibits very good performance: in a closed-world setting with 100 sites, it achieved 91\% accuracy, and in an open-world setting with 5,000 sites, it achieved 86\% True Positive Rate (TPR) and 0.6\% False Positive Rate (FPR).

\paragraphX{CUMUL.}
Panchenko et al.~\cite{Panchenko2016} proposed an attack based on a Support Vector Machine (SVM) classifier and devised a novel feature set based on the cumulative sum of packet lengths constructed as follows: the first coordinate in the feature vector is the length of the first packet in the traffic trace and the $i$-th coordinate is the sum of the value in the ($i-1$)-th coordinate plus the length of the $i$-th packet, where lengths for incoming packets are negative. The attack achieved 91\% accuracy in a closed-world setting. In the open-world, they study two different scenarios: \textit{multi-class}, where each monitored site is treated as a different class, and \textit{two-class}, where the whole set of monitored pages is treated as a single class. The open world results are 96\% TPR and 9.61\% FPR for \textit{multi-class} and 96\% TPR and 1.9\% FPR for {two-class}.


\paragraphX{$k$-FP.}
Hayes and Danezis~\cite{Hayes2016} proposed the \textit{k}-fingerprinting attack ($k$-FP). $k$-FP uses a random forest classifier to extract fingerprints of pages: they train the random forest with traditional features, but the actual fingerprint is represented by the leafs of the trees in the random forest. The authors argue this representation is more effective for WF than the one based on the original features.
To solve the open world problem, they feed these new feature vectors to a $k$-NN classifier. They also analyze the importance of their features and ranked them. The results show that the top 20 most important features involve counting the number of packets in a sequence, and that these leak more information about the identity of a web page than complex features such as packet ordering or packet inter-arrival time features. $k$-FP achieved 91\% accuracy in a closed-world setting and 88\% TPR and a 0.5\% FPR in an open-world setting.  

\subsection{WF defenses}
The fundamental strategy to defend against WF attacks is to add dummy packets and/or delay packets. This \emph{cover traffic} makes WF features less distinctive, thus increasing the rate of classification errors committed by the adversary. The first defense that used this strategy against WF was BuFLO~\cite{Dyer2012}, proposed by Dyer et al., whose strategy was to modify the traffic to make it look constant rate and thus remove packet-specific features. However, coarse features such as total volume, size and time were hard to conceal without incurring high bandwidth overheads~\cite{Dyer2012}. Tamaraw~\cite{Cai2014a} and CS-BuFLO~\cite{Cai2014b} tried to solve this problem by grouping sites that are similar in size and padding all the sites in a group to the greatest size in that group. Even so, these defenses still require more than 130\% extra bandwidth than unprotected Tor and, on average, pages load between two to four times slower~\cite{Dyer2012,Cai2014a,Cai2014b}.

Recently, two lightweight countermeasures have been proposed for deployment in Tor for their low latency overhead: WTF-PAD and Walkie-Talkie.

\paragraphX{WTF-PAD.}
Tor developers have expressed a preference for using {\em adaptive padding} as a WF defense~\cite{Perry2013,Perry2015}. Adaptive padding~\cite{Shmatikov2006} saves bandwidth by adding the padding only upon low usage of the channel, thus masking traffic bursts and their corresponding features. Since adaptive padding was originally designed as a defense against end-to-end timing analysis, Juarez et al.\ proposed WTF-PAD, a system design for deploying adaptive padding for WF defense in Tor~\cite{JR2016}. WTF-PAD has been shown to be effective against all state-of-the-art attacks with relatively moderate bandwidth overheads compared to the BuFLO-style defenses (e.g. 54\%). Plus, since WTF-PAD does not delay packets, it does not incur any latency overhead. 

\paragraphX{Walkie-Talkie.}
Walkie-Talkie (W-T) has the Tor browser communicate with the web server in \textit{half-duplex} mode, in which the client sends a request (such as for an image file) only after the server has fulfilled all previous requests. As a result, the server and the client send non-overlapping bursts in alternate directions. Moreover, the defense also adds dummy packets and delays to create {\em collisions}, in which two or more sites have the same features as used by the adversary's classifier. The key idea is that the traces that result from half-duplex communication can be transformed to create a collision with less padding than it would with full-duplex traces. W-T provides strong security guarantees with 31\% bandwidth overhead and 34\% latency overhead.

Despite the low cost of these two defenses, their evaluations have shown that each defense can significantly reduce the accuracy of the attacks to less than 30\%. As of today, they are the main candidates to be implemented in Tor. In this paper, we evaluate all the attacks against them.

\subsection{WF Attacks using Deep Learning}
Many applications have adopted deep learning (DL) to solve complex problems such as speech recognition, visual object recognition, and object detection in images~\cite{LeCun2015}. DL does not require selecting and fine-tuning features by hand. In the WF domain, there are four works that have begun to examine the use of DL. 

Abe and Goto studied the application of Stacked Denoising Autoencoders (SDAE)~\cite{Abe2016} to WF attacks. They showed that SDAE is effective with 88\% accuracy in the closed world and 86\% TPR and 2\% FPR in the open world. Although most work in deep learning recommends large data sets be used, their work was successful with only a small dataset. 

Rimmer et al.\ proposed to apply DL for automated feature extraction in WF attacks~\cite{Rimmer2018}. The results show that the adversary can use DL to automate the feature engineering process to effectively create WF classifiers. Thus, it can eliminate the need for feature design and selection. In the closed-world scenario, their CNN-based attack (which we refer to as {\em Automated Website Fingerprinting}, or AWF) trained on 2,500 traces per site could achieve 96.3\% accuracy. In their open-world evaluation, SDAE performs the best of their models with 71.3\% TPR and 3.4\% FPR when optimizing for low FPR. However, AWF could not outperform state-of-the-art WF attacks such as CUMUL. 

Recently, Bhat et al.~\cite{Bhat2018} and Oh et al.~\cite{Oh2018} have released preliminary reports on their explorations of a CNN variant and unsupervised DNNs with autoencoders, respectively. While both papers include interesting contributions, neither paper reports accuracy rates as high as those shown in our results. Additionally, neither attack was shown to be effective against WTF-PAD.




In this work we aim to bridge this gap by developing a powerful CNN-based deep learning model called deep fingerprinting (DF) that can substantially outperform all previous state-of-the art WF attacks. The DF model uses a more sophisticated variant of CNN than AWF, with more convolutional layers, better protections against overfitting, hyperparameters that vary with the depth of each layer, activation functions tailored to our input format, and a two-layer fully connected classification network. These differences in the architectural model from AWF, which are described in more detail in Section~\ref{IntellectualDiff}, lead to a deeper and more effective network. We show that the DF model works significantly better than AWF and all other attacks, particularly against WF defenses and in the more realistic open-world setting.


\subsection{Deep Learning}

In our work, we mainly focus on two deep learning techniques that previous work has shown to be promising for WF attacks.

\begin{figure*}[t]
  \centering
  \includegraphics[width=0.78\textwidth]{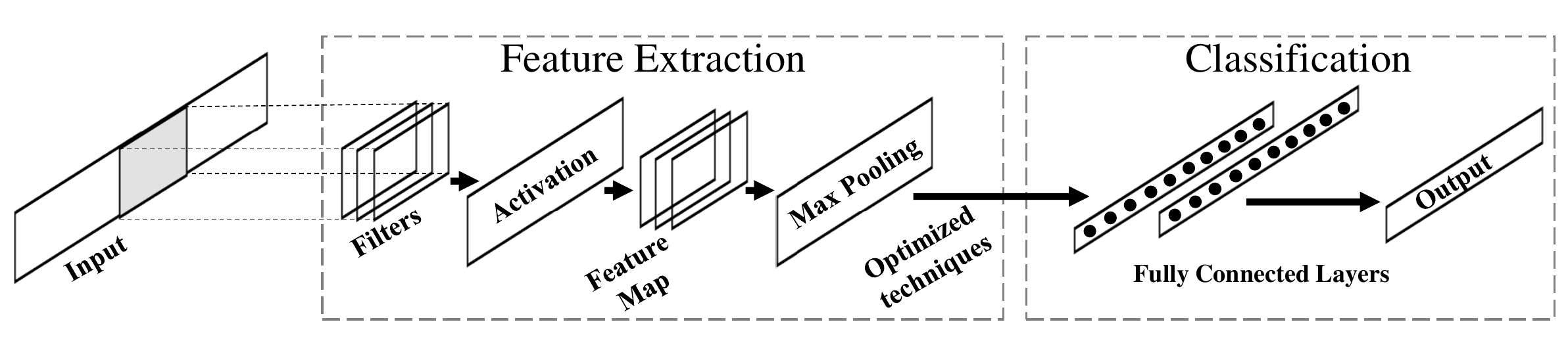}
  \caption{A basic architecture of convolutional neural networks (CNN)}\label{fig:CNN}
\end{figure*}

\subsubsection{Stacked Denoising Autoencoders (SDAE)}
Vincent et al.~\cite{Vincent2010} proposed SDAEs in 2010 to improve classification performance in recognizing visual data. SDAE leverages the concept of an autoencoder (AE), a simple 3-layer neural network including input, hidden and output layers. In AE, the input data is first {\em encoded}, passing it through a layer of neurons to a more condensed representation (the hidden layer). The AE then performs {\em decoding}, in which it attempts to reconstruct the original input from the hidden layer while minimizing error. 
The main benefit of AE is to extract high-level features from the training data, resulting in dimensionality reduction.  

A Denoising Autoencoder (DAE) uses the basic concept of AE but also adds noise to the input. 
The DAE tries to reconstruct the original values from the noisy inputs, which helps it to better generalize and thus handle a wider variety of inputs after training. 
SDAE combines ("stacks") multiple DAEs by overlapping a hidden layer as an input of the next DAE. 
Vincent et al.\ showed that SDAE achieves lower classification error rates for image classification compared to SVM, Deep Belief Networks (DBN), and Stacked Autoencoders (SAE)~\cite{Vincent2010}.

\subsubsection{Convolutional Neural Networks (CNN)}
CNNs have become the gold standard in image classification after Krizhevsky et al.\ won the Large Scale Visual Recognition Challenge (ILSVRC) in 2012~\cite{Krizhevsky2012}. 
Schuster et al.\ recently proposed applying a CNN on encrypted video streams, and they show that the encrypted stream could be uniquely characterized by their burst patterns with high accuracy~\cite{Schuster2017}. This suggests that CNNs could be useful for WF attacks as well. Figure~\ref{fig:CNN} shows the basic architecture of a CNN~\cite{Lecun1998,Krizhevsky2012}. The architecture consists of two major components: \textit{Feature Extraction} and \textit{Classification}.

In \textit{Feature Extraction}, the input is first fed into a \textit{convolutional layer}, which comprises a set of {\em filters}. Each region of input is {\em convolved} with each filter, essentially by taking the dot product of the two vectors, to get an intermediate set of values. These values are input to an {\em activation function} -- this is similar to neurons being activated based on whether or not the filtered input has certain features. Having more filters means being able to extract more features from the input. The output of the activation function is then fed into a \textit{pooling} layer.
The pooling layer progressively reduces the spatial size of the representation from the feature map to reduce the number of parameters and amount of computation. The most common approach used in pooling is \textit{Max Pooling}, which simply selects the maximum value in a spatial neighborhood within a particular region of the feature map to be a representation of the data. 
This has the advantage of being invariant to small transformations, distortions and translations in the input, since the largest signals in each neighborhood are retained.
The final part of the feature extraction component (\textit{Optimized techniques} in Figure~\ref{fig:CNN}) mainly consists of a stochastic \textit{dropout} function  and \textit{Batch Normalization} that help improve classifier performance and prevent overfitting. 

The CNN then passes the output from the convolutional and pooling layers, which represents high-level features of the input, into the \textit{Classification} component. In this component, a set of fully-connected layers uses the features to classify the input. During training, the loss value of classification is used to not only update weights in the classification component but also the filters used in feature extraction. To estimate the loss value, we use \textit{categorical cross-entropy}, which is suitable for multi-class classification problems such as WF.


\section{Data Collection}\label{sec:data_collection}

For the closed-world dataset, we visited the homepage of each of the top Alexa 100 sites 1,250 times and dumped the traffic generated by each visit separately using \texttt{tcpdump}. We used ten low-end machines in our university's campus to collect the data. We have followed prior work's methodology for data collection~\cite{Wang2013,Juarez2014}; on each machine, the visits were sequential and were ordered according to Wang and Goldberg's batched methodology to control for long- and short-term time variance~\cite{Wang2013}.
More specifically, we split the visits to each site in five chunks, so that the websites are accessed in a round-robin fashion: in each batch we access \emph{each} site 25 times. As a result of batching, the visits to a site are spread over time. The rationale for this is twofold: i) to avoid having our IP addresses banned by the web servers; and, ii) to capture variants of the sites over time for more robust training and testing.

We used \texttt{tor-browser-crawler}~\cite{Juarez2014} to drive the Tor Browser to visit websites. This allows for more realistic crawls than using tools like \texttt{wget} or \texttt{curl} because the setting resembles a real user browsing the Web with Tor. We acknowledge that to be more realistic, our crawler should model user browsing behavior when crawling sites. However, modeling user behavior in Tor is challenging, as user statistics are not collected for privacy reasons. Virtually all existing datasets collected for WF follow the same simplistic user model we use in this study.

After the crawls were finished, we discarded corrupted traffic traces. For instance, we removed traces that did not have any incoming or outgoing packets or were too short -- less than 50 packets. After removing corrupted traces, we only kept the sites, or classes, that had at least 1,000 visits. We ended having 95 sites with 1,000 visits for our closed-world evaluations. We refer to the set of the data used for closed-world evaluations as the \textit{closed-world} dataset.

\paragraphX{Open-world dataset.}
For the open-world dataset, we visited the sites from Alexa's top 50,000, excluding the first 100 sites used to build the closed-world dataset. We used the same ten machines to collect the data, where each machine collected the data for 5,000 different sites sequentially. We visited each open-world site only once and took a screenshot of their homepages. After collecting the data, we discarded corrupted visits the same way we did for the closed-world dataset. During the crawling of the open-world, we found sites returning an access denied error message, a timeout error, or a blank page. Moreover, many of the sites were behind Cloudflare's CDN, which presents a CAPTCHA to connections coming from Tor exit relays. We removed those sites from our dataset by comparing their homepage's screenshot with each of: a blank page, an access denied page, a CAPTCHA page, and a timeout error page. The final dataset has a total of 40,716 traffic traces. 

\paragraphX{Defended dataset.}
To evaluate the defenses, we produced datasets with traces protected by each defense: for BuFLO, Tamaraw and WTF-PAD, we protect traces by padding them according to the defense protocols, using the scripts and simulators provided by the authors~\cite{Dyer2012,Cai2014a,JR2016}. 
Walkie-Talkie, however, cannot be completely simulated, as half-duplex communication is hard to model. We thus performed a new crawl with a Tor Browser in half-duplex mode. Since the implementation of half-duplex for the original implementation of Walkie-Talkie was done in an outdated version of the Tor Browser, we had to implement half-duplex in the latest version of Tor Browser at the time of our crawls (Tor Browser Bundle version 7.0.6). With this modified Tor Browser, we collected closed- and open-world datasets of size similar to the undefended ones. Walkie-Talkie also requires padding the bursts in the half-duplex traces. To that end, we followed the \emph{mold} padding strategy as described in the Walkie-Talkie paper~\cite{Wang2017}.

\begin{table}[tbp]
\small
\centering
\caption{Hyperparameters selection for DF model from Extensive Candidates Search method}
\label{table:hyperparameters}
\begin{tabular}{@{}lll@{}}

\toprule
\textbf{Hyperparameters}                                                                                                                                                                  & \textbf{Search Range}                                                                                            & \textbf{Final}                                                                                        \\ \midrule
Input Dimension                                                                                                                                                                           & {[}500 ... 7000{]}                                                                                               & 5000                                                                                                  \\
Optimizer                                                                                                                                                                                 & \begin{tabular}[c]{@{}l@{}}{[}Adam, Adamax, \\ RMSProp, SGD{]}\end{tabular}                                      & Adamax                                                                                                \\
Learning Rate                                                                                                                                                                             & {[}0.001 ... 0.01{]}                                                                                             & 0.002                                                                                                 \\
Training Epochs                                                                                                                                                                           & {[}10 ... 50{]}                                                                                                  & 30                                                                                                    \\
Mini-batch Size                                                                                                                                                                           & {[}16 ... 256{]}                                                                                                 & 128                                                                                                   \\
{[}Filter, Pool, Stride{]} Sizes                                                                                                                                                           & {[}2 ... 16{]}                                                                                                   & {[}8, 8, 4{]}                                                                                         \\
Activation Functions                                                                                                                                                                      & {[}Tanh, ReLU, ELU{]}                                                                                            & ELU, ReLU                                                                                             \\
\begin{tabular}[c]{@{}l@{}}Number of Filters\\    ~~~Block 1 {[}Conv1, Conv2{]}\\    ~~~Block 2 {[}Conv3, Conv4{]}\\    ~~~Block 3 {[}Conv5, Conv6{]}\\    ~~~Block 4 {[}Conv7, Conv8{]}\end{tabular} & \begin{tabular}[c]{@{}l@{}}\\{[}8 ... 64{]}\\ {[}32 ... 128{]}\\ {[}64 ... 256{]}\\ {[}128 ... 512{]}\end{tabular} & \begin{tabular}[c]{@{}l@{}}\\{[}32, 32{]}\\ {[}64, 64{]}\\ {[}128, 128{]}\\ {[}256, 256{]}\end{tabular} \\
Pooling Layers                                                                                                                                                                            & {[}Average, Max{]}                                                                                               & Max                                                                                                   \\
Number of FC Layers                                                                                                                                                                       & {[}1 ... 4{]}                                                                                                    & 2                                                                                                     \\
Hidden units (each FCs)                                                                                                                                                                   & {[}256 ... 2048{]}                                                                                               & {[}512, 512{]}                                                                                        \\
Dropout {[}Pooling, FC1, FC2{]}                                                                                                                                                                   & {[}0.1 .. 0.8{]}                                                                                                 & {[}0.1, 0.7, 0.5{]}                                                                                        \\ \bottomrule
\end{tabular}
\end{table}

\section{Experimental Evaluation}\label{sec:Evaluation}
						In this section, we evaluate WF attacks based on SDAE, DF and AWF. We compare them with the state-of-the-art WF attacks. We used our datasets for these evaluations.

\subsection{Implementation}
Our implementation of the DF model uses the Python deep learning libraries \textit{Keras} as the front-end and \textit{Tensorflow} as the back-end~\cite{keras}. The source code of the implementation and a dataset to reproduce our results is publicly available at~\url{https://github.com/deep-fingerprinting/df}.

\subsubsection{Data Representation}\label{datarep}
In WF, a website trace is represented as a sequence of tuples $<$$timestamp$, $\pm packet\_size$$>$, where the sign of $packet\_size$ indicates the direction of the packet: positive means outgoing and, negative, incoming.

Prior work in WF has shown that the most important features are derived from the lengths of traces in each direction~\cite{Wang2014,Hayes2016}. Wang et al.~\cite{Wang2014} simplified the raw traffic traces into a sequence of values from [$-1$, $+1$], where they ignored packet size and timestamps and only take the traffic direction of each packet. However, we performed preliminary evaluations to compare the WF attack performance between using packet lengths and without packet lengths, i.e., only packet direction, as feature representations. Our result showed that using packet lengths does not provide a noticeable improvement in the accuracy of the attack. Therefore, we follow Wang et al.'s methodology and consider only the direction of the packets.


SDAE, DF and AWF require the input to have a fixed length. In order to find the input length that performs best, we parameterized it and explored the range $[500,\hspace{0.1cm}7,000]$, which contains most of the length distribution in our data. Our results show that 5,000 cells provide the best results in terms of classification accuracy. In practice, most of the traces are either longer or shorter than that. We padded shorter traces by appending zeros to them and truncated longer traces after 5,000 cells. Out of 95,000 traces in the closed-world dataset, only 8,121 were longer than 5,000 cells and had to be truncated, while the rest were padded.


\subsubsection{SDAE}
We reproduced Abe and Goto's results~\cite{Abe2016}, as described in Section~\ref{sec:related}.
Following guidance from the authors, we successfully re-implemented their SDAE neural network on the same architecture and dataset they used. We achieved 89\% accuracy, a slightly higher accuracy than the one Abe and Goto reported in their paper. We believe that using different Python DL modules (\textit{Kera} and \textit{Tensorflow}) and randomly initializing the weights accounts for this difference. Furthermore, we slightly changed their SDAE model's hyperparameters to improve its performance in our experiments. 

\subsubsection{AWF}
Rimmer et al.\ provided us with their source code
to re-produce their results.
We strictly followed their proposed hyperparameters and evaluate the model in our dataset to make a fair comparison for our model and the previous state-of-the-art WF attacks.

\subsubsection{DF}
\label{DF}
To develop our DF model to effectively perform WF attacks on both non-defended and defended dataset, we have followed techniques in the deep learning literature~\cite{Krizhevsky2012, Simonyan2015, Szegedy2015} to improve the performance of the model, such as using the appropriate number of convolutional layers, mitigation and prevention of overfitting~\cite{Srivastava2014} and a suitable activation function for our WF input data~\cite{Mishkin2016}. These studies helped us design the sophisticate architecture and tune the model's hyperparameters that best fit for WF.

We adapted the base CNN model of DF to our needs, as there are important differences between traffic analysis and traditional applications of CNN-based models such as image recognition. For example, standard activation functions such as \textit{sigmoid} and {\em rectified linear unit (ReLU)} do not activate on negative values and thus will not use the information conveyed by the sign of the input (i.e., cell direction). Activation functions that can handle negative inputs include tanh, leaky ReLU (LReLU), parametrized ReLU (PReLU) and Exponential Linear Unit (ELU). Prior work has shown that ELU provides fast and accurate classification~\cite{Clevert2015,Mishkin2016}. We compared ELU with other activation functions during hyperparameter tuning and it performed the best among all the functions we tested (see Section~\ref{hyper}).
Although the traditional $\tanh$ function can also handle negative inputs, it is vulnerable to the \textit{vanishing gradient} issue~\cite{Bengio1994}, which slows down optimization.

Another difference between traffic and image data is that images are two-dimensional, whereas our data is a vector. This means that filter and pool sizes cannot be two-dimensional (e.g., 2x4) but have to be cast to one dimension (e.g., 1x4).





\subsection{DF's Hyperparameter Tuning}
\label{hyper}
A fundamental process in supervised classification is to tune the {\em hyperparameters} of the classification model, such as the kernel used in an SVM or the number of hidden layers in a CNN. This process involves adjusting the trade-off between variance, bias and classification performance, so that the model fits the training data while still generalizing to samples that it has not been trained on. 
For DF, however, the large amount of training data and the large number of hyperparameters the model has render an exhaustive search prohibitive in terms of computational resources. To demonstrate our attacks, we thus only aim at a good-enough classifier and we acknowledge that someone with more resources might be able to optimize our model further.

To select the hyperparameters for our models, we perform an extensive search through the hyperparameter space, in which we build each layer of the deep learning model block by block. In each building block, we vary the hyperparameters to estimate the gradient of the parameter and determine whether we must increase or decrease its value. Once this process is done, we select the best top-\textit{n} parameters and use them as the initial parameters for the optimization in the next block. When all layers are set, we select the best combination of hyperparameters.

By the transferability property of neural networks~\cite{Yosinski2014}, WF attacks based on other models can use the values we found for the DF hyperparamters to bootstrap the hyperparameters of their model. We thus used the transferability property to find the parameters for the defended datasets from the hyperparameters found using the undefended dataset. We only needed to slightly adjust some hyperparameters to optimize our model, significantly reducing the time spent in hyperparameter tuning. We thoroughly illustrate and explain our design of DF model in Appendix~\ref{appendix:CNNModel}. The search space as well as the final selected values are shown in Table~\ref{table:hyperparameters}. 

\paragraphX{Evaluating overfitting}
Even though deep neural networks (DNN) are a powerful supervised classification model, they are, as with as most machine learning models, vulnerable to overfitting. Overfitting occurs when the model errs on samples that it has not been trained on, meaning that the model cannot generalize. For small datasets, overfitting can be measured with cross-validation techniques. For large datasets like ours, however, we can just split the data into three mutually exclusive sets: training, validation and testing, with a ratio of 8:1:1. We then measure the difference in error rate between different sets. In case of overfitting, the model would achieve a substantially higher accuracy on the training set than on the testing set.

During the training of our DF model, we applied \textit{Dropout}~\cite{Srivastava2014} and \textit{Batch Normalization (BN)}~\cite{Ioffe2015} to prevent overfitting. These are regularization techniques that loosen the model and allow for greater generalization. In dropout, the model randomly selects hidden units, including their incoming and outgoing connections, and temporarily removed them from the network while training. BN normalizes the fully-connected and convolutional layers' outputs and helps accelerate learning while also reducing overfitting. Moreover, we analyze the error rates between training and testing datasets during hyperparameter tuning to ensure that our model is not overfitting.

Figure~\ref{fig:traing_epochs} depicts the training and testing error rates. The difference between training and testing error of the DF model is less than 2\%, suggesting that overfitting is unlikely.

\subsection{Differences Between DF and AWF}
\label{IntellectualDiff}
We now explain the significant differences of the DF model compared to the AWF model proposed by Rimmer et al.~\cite{Rimmer2018} that help explain the superior performance of DF.

\paragraphX{Basic Block Design.}
The basic block is the group of convolutional layer(s), max pooling layer, filters and activation layer(s) that perform feature extraction in a CNN. Generally, the basic block is repeatedly appended to create deeper networks.

\begin{figure}[t]
  \centering
  \includegraphics[width=0.48\textwidth]{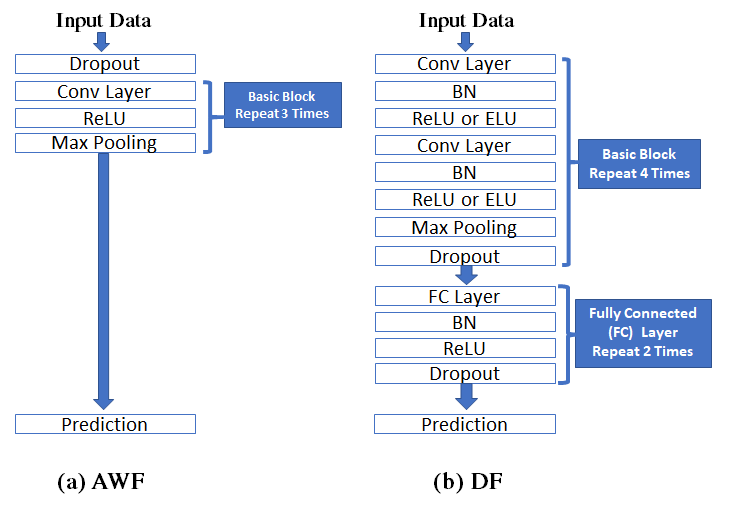}
  \caption{Comparison between AWF and DF models}\label{fig:AWFvsDF}
\end{figure}

We observe that the AWF model is similar to Imagenet~\cite{Krizhevsky2012}, one of the earliest CNN models proposed in 2012. The basic block of this model only contains one convolutional layer followed by one max pooling layer as shown in Figure~\ref{fig:AWFvsDF}(a). In contrast, our DF model is inspired by modern large image classification networks such as VGG~\cite{Simonyan2015}, GoogleNet~\cite{Szegedy2015} and ResNet~\cite{he2016} that apply at least two consecutive convolutional layers before a max pooling layer as shown in Figure~\ref{fig:AWFvsDF}(b). Max pooling typically reduces the data to a smaller size, so it is not possible to have deeper networks when pooling after every convolutional layer. Adding more convolutional layers in each basic block thus enables more convolutional layers in total and a deeper network with more effective feature extraction.

\paragraphX{Overfitting Concerns.}
Rimmer et al.\ criticized the CNN model for having a higher risk of overfitting which was shown in their experimental results. We argue that a more carefully crafted model can mitigate overfitting. The AWF model includes a dropout layer before the first basic block as shown in Figure~\ref{fig:AWFvsDF}(a). While dropout is a common technique to help prevent overfitting, this placement is atypical in CNN designs, as it may result in the loss of meaningful features extracted from the input and may not be sufficient to prevent overfitting. In DF, we used overfitting-contention mechanisms that are applied in the state-of-the-art CNN networks for computer vision, including a batch normalization (BN) layer that is added right after each convolutional layer and a dropout layer  after the activation function, as explained in Section~\ref{hyper}. With these mechanisms, the DF model shows no evidence of overfitting in our experiments.


\paragraphX{Varying Hyperparameters}
In the AWF model, the value of some hyperparameters are fixed such as using 32 filters in every convolutional layer. Using a fixed number of filters over all the layers reduces the capability of the model to learn. In contrast, the DF model follows the state-of-the-art in computer vision by varying hyperparameter values for different layers~\cite{Simonyan2015}. For example, we increase the number of filters as we get deeper in the network. The intuition behind varying the values is that the CNN uses hierarchical features in its processing pipeline. The features in lower layers (close to the input) are primitive, like edge detection, while features in upper layers are high-level abstract features, like object detection, made from combinations of lower-level features. We increase the number of filters at higher layers to improve the ability to encode richer representations.

\paragraphX{Activation Function.}
The AWF model only uses the ReLU activation function. ReLU is popular in CNNs, but it maps all negative values to zero. Our input formats include negative values that represent incoming packets, so using ReLU in convolutional layers close the input can substantially reduce the information available to deeper layers in the model. In the DF model, the activation function in the first basic block is ELU, which can learn a data representation containing negative values to ensure that the model can learn all meaningful representations from the input.


\paragraphX{Fully-connected Layers.}
The AWF model directly connects the last max pooling layer to the prediction layer, a densely connected layer with an output size equal to number of classes. In more recent CNNs, there are a set of fully connected (FC) layers that follow the convolutional layers and precede the prediction layer. The FC layers play an important role in the learning and classification processes of the model. Essentially, the convolutional layers perform feature extraction, but that means that it is important to carefully design the classifier that uses the extracted features, which is the role of the FC layer. In the DF model, we add two FC layers with the combination of BN and dropout to prevent the overfitting that normally occurs in FC layers. 



\vspace{0.2cm}
Overall, our DF model was specifically designed to effectively perform WF attacks by leveraging the state-of-the-art techniques from computer vision research. We  provide a thorough explanation on how the DF model was developed and a visualization of the DF model in Appendix~\ref{appendix:CNNModel} to allow other researchers to gain better understanding and reproduce our work. Our experimental results confirm that DF performs better than AWF model in defended and non-defended and on both closed-world and more realistic open-world scenarios. These results help to illustrate the impact of the DL architecture design on the performance of the attacks. 

\subsection{Closed-world Evaluation on Non-defended Dataset}
We evaluate the performance of the DF attack in the closed-world scenario on the \textit{non-defended} dataset, which comprises website traces from the \textit{closed-world} dataset with no WF defenses in place. Moreover, we compare DF model with the state-of-the-art WF attacks: $k$-NN, \textit{CUMUL}, $k$-FP, AWF, and SDAE. We re-evaluate these attacks on our \textit{non-defended} dataset and apply \textit{$k$-fold cross-validation} for training their classifiers and testing their performance, as was done in the papers presenting these attacks~\cite{Wang2014,Panchenko2016,Hayes2016}. 

\begin{table}[hbp]
\renewcommand{\arraystretch}{1.2}
\small
\centering
\caption{{\bf Closed World:} Accuracy on the non-defended dataset for state-of-the-art attacks.}
\label{accuracy-close-world-nondef}
\begin{tabular}{c|p{0.8cm}p{0.7cm}p{0.7cm}p{0.8cm}p{1.0cm}p{0.8cm}}
\hline
\textbf{Classifier} & \textbf{SDAE} & \textbf{DF} & \textbf{AWF} & \textbf{\textit{$k$-NN}} & \textbf{CUMUL} & \textbf{\textit{$k$-FP}} \\ \hline
\textbf{Accuracy}   & 92.3\%       & \textbf{98.3}\%  & 94.9\%    & 95.0\%       & 97.3\%        & 95.5\%       \\ \hline
\end{tabular}
\end{table}

Table~\ref{accuracy-close-world-nondef} shows the accuracy results. Our DF model attains 98.3\% accuracy, which is better than the other attacks and higher than any previously reported result for a WF attack in Tor. Our DF model performs better than AWF. Our results for AWF (94.9\%) are a bit lower than the reported results by Rimmer et al.\ (96.5\%), we believe this is due to the larger dataset used by them. We observe that CUMUL, $k$-FP, $k$-NN and SDAE benefit from the larger training data set with 2-4\% higher accuracies than previously reported results that used smaller datasets (usually 90 training instances). SDAE was not as accurate as the other attacks. 

\begin{figure}[t]
  \centering  
\begin{tikzpicture}

\definecolor{color1}{rgb}{0.75,0.75,0}
\definecolor{color0}{rgb}{0.75,0,0.75}

\begin{axis}[
width=210,
height=170,
xlabel={Number of epochs},
ylabel={Accuracy},
xmin=8, xmax=42,
ymin=90, ymax=100,
axis on top,
xtick pos=both,
ytick pos=left,
xmajorgrids,
x grid style={lightgray!66.928104575163388!black},
ymajorgrids,
y grid style={lightgray!66.928104575163388!black}
]
\addplot [thick, blue, dashed, mark=square*, mark size=\MarkerSize, mark options={solid}, line width=\LineWidth]
table {%
10 98.58
15 99.02
20 99.24
25 99.48
30 99.55
35 99.57
40 99.62
};
\addplot [thick, red, dashed, mark=triangle*, mark size=\MarkerSize, mark options={solid,rotate=180}, line width=\LineWidth]
table {%
10 97.61
15 97.76
20 97.86
25 98.27
30 98.25
35 98.44
40 98.29
};
\end{axis}

\begin{axis}[
width=210,
height=170,
ylabel={Error Rate},
xmin=8, xmax=42,
ymin=0, ymax=10,
axis on top,
axis y line=right,
xtick pos=both,
ytick pos=right,
legend style={at={(0.95,0.5)}, anchor=east},
legend style={font=\scriptsize},
legend cell align={left},
legend entries={{DF Training Accuracy},{DF Testing Accuracy},{DF Testing Error Rate},{DF Training Error Rate}}
]
\addlegendimage{mark=square*, blue, dashed, mark size=2}
\addlegendimage{mark=triangle*, red, dashed, mark options={solid,rotate=180}, mark size=2}
\addlegendimage{mark=otimes*, color0, dashed, mark size=2}
\addlegendimage{mark=diamond*, green!50.0!black, dashed, mark size=2}
\addplot [thick, dashed, color0, mark=otimes*, mark size=\MarkerSize, mark options={solid}, line width=\LineWidth]
table {%
10 2.39
15 2.23999999999999
20 2.14
25 1.73
30 1.75
35 1.56
40 1.70999999999999
};
\addplot [thick, dashed, green!50.0!black, mark=diamond*, mark size=\MarkerSize, mark options={solid}, line width=\LineWidth]
table {%
10 1.42
15 0.980000000000004
20 0.760000000000005
25 0.519999999999996
30 0.450000000000003
35 0.430000000000007
40 0.379999999999995
};
\end{axis}

\end{tikzpicture}
  \caption{{\bf Closed World:} Impact of the number of training epochs on DF accuracy and error rate}\label{fig:traing_epochs}
\end{figure}
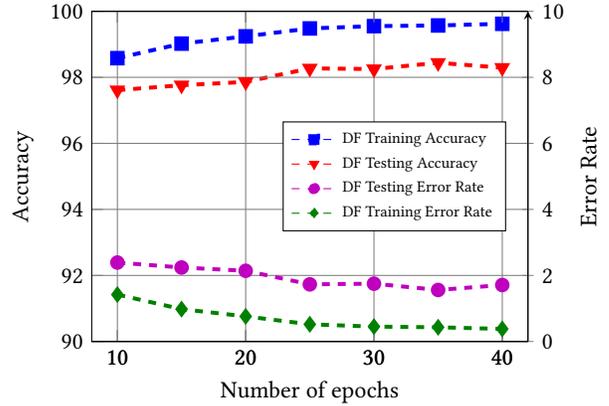

\begin{figure}[t]
  \centering
\begin{tikzpicture}

\definecolor{color1}{rgb}{0,0.75,0.75}
\definecolor{color0}{rgb}{0.75,0,0.75}
\definecolor{color0}{rgb}{0.75,0,0.75}

\begin{axis}[
width=220,
height=220,
xlabel={Training Size},
ylabel={Accuracy},
xmin=0, xmax=950,
ymin=60, ymax=100,
axis on top,
tick pos=both,
xmajorgrids,
x grid style={lightgray!60.928104575163388!black},
ymajorgrids,
y grid style={lightgray!60.928104575163388!black},
legend entries={{DF},{CUMUL},{k-NN},{k-FP},{AWF},{SDAE}},
legend cell align={left},
legend style={at={(0.97,0.03)}, anchor=south east}
]
\addlegendimage{dashed, mark=otimes*, mark size=2, mark options={solid}, line width=1.0, blue}
\addlegendimage{dashed, mark=triangle*, mark size=2, mark options={solid,rotate=180}, line width=1.0, green!50.0!black}
\addlegendimage{dashed, mark=diamond*, mark size=2, mark options={solid}, line width=1.0, red}
\addlegendimage{dashed, mark=triangle*, mark size=2, mark options={solid}, line width=1.0, color1}
\addlegendimage{dashed, mark=halfcircle*, mark size=2, mark options={solid, rotate=45}, line width=1.0, brown!50.0!black}
\addlegendimage{dashed, mark=square*, mark size=2, mark options={solid}, line width=1.0, color0}
\addplot [thick, blue, dashed, mark=otimes*, mark size=2, mark options={solid}, line width=1.0]
table {%
50 90.99
100 93.76
250 96.85
350 97.26
550 97.85
650 98.04
750 98.13
900 98.44
};
\addplot [thick, green!50.0!black, dashed, mark=triangle*, mark size=2, mark options={solid,rotate=180}, line width=1.0]
table {%
50 90.78
100 93.17
250 95.39
350 96.39
550 96.28
650 96.77
750 97
900 97.26
};
\addplot [thick, red, dashed, mark=diamond*, mark size=2, mark options={solid}, line width=1.0]
table {%
50 85.95
100 89.35
250 92.48
350 93.17
550 93.68
650 94.15
750 94.88
900 95
};
\addplot [thick, color1, dashed, mark=triangle*, mark size=2, mark options={solid}, line width=1.0]
table {%
50 84.74
100 88.29
250 91.59
350 92.18
550 93.37
650 93.66
750 94.08
900 94.5
};
\addplot [thick, brown!50.0!black, dashed, mark=halfcircle*, mark size=2, mark options={solid,rotate=45}, line width=1.0]
table {%
50 76.56
100 84.68
250 91.41
350 92.78
550 93.73
650 93.98
750 94.65
900 94.91
};
\addplot [thick, color0, dashed, mark=square*, mark size=2, mark options={solid}, line width=1.0]
table {%
50 62.85
100 75.04
250 83.56
350 86.34
550 88.25
650 89.84
750 90.16
900 92.34
};
\end{axis}

\end{tikzpicture}
  \caption{{\bf Closed World:} Impact of numbers of training traces on classification accuracy}\label{fig:Growing}
\end{figure}
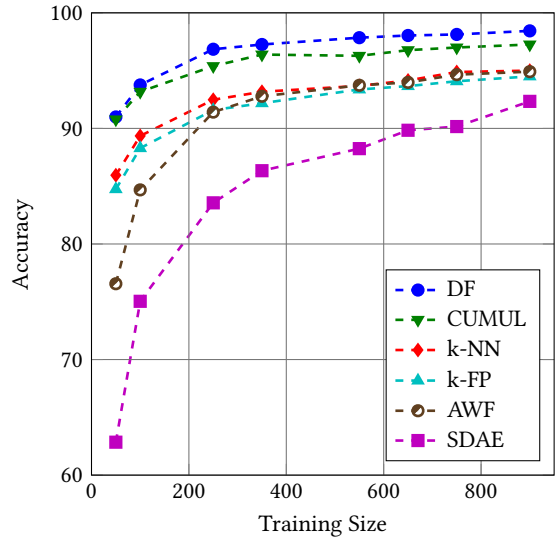

Additionally, we investigate how fast the model can learn to distinguish patterns from the input data, also known as \emph{convergence} of the model. This depends on many factors such as the method used for data pre-processing, the DL architecture and the hyperparameters used to create the model. This evaluation helps to validate the quality of our hyperparameter tuning method. Moreover, the attacker can use this to estimate the number of {\em training epochs}, rounds of feeding training data into the classifier, required for the classifier to reach the expected level of accuracy. Normally, the classifier gradually learns better with more training epochs. 

Figure~\ref{fig:traing_epochs} shows that with only 10 training epochs, DF can reach testing accuracy of about 97\%, DF consistently improves with more training epochs, until accuracy levels off after 30 epochs. 

Finally, we investigate the impact of dataset size on classifier accuracy. The results shown in Figure~\ref{fig:Growing} indicate that DF and CUMUL consistently outperform the other attacks for all training sizes. With just 50 traces per site, both DF and CUMUL achieve 90\% accuracy. $k$-NN, $k$-FP and AWF require 250 traces to reach this accuracy, and SDAE requires 750 traces. The observed accuracies mostly saturate after 550 traces, except for SDAE. The results show that the various techniques used in the DF model lead to significantly better performance compared to the simpler AWF model.   

\begin{table*}[tbp]
\small
\renewcommand{\arraystretch}{1.25}
\centering
\caption{Accuracy in a closed-world scenario on defended datasets, SDAE, DF, and AWF vs. the state-of-art WF attacks}
\label{accuracy-close-world-def}
\begin{tabular}{c|cc|cccccc}
\hline
\multirow{2}{*}{\textbf{Defenses}} & \multicolumn{2}{c|}{\textbf{Overhead}} & \multicolumn{6}{c}{\textbf{Accuracy of WF attacks on defended datasets}}           \\ \cline{2-9} 
                                   & \textbf{Bandwidth}  & \textbf{Latency}  & \textbf{SDAE} & \textbf{DF} & \textbf{AWF}   & \textbf{\textit{$k$-NN}} & \textbf{CUMUL} & \textbf{\textit{$k$-FP}} \\ \hline
\textbf{BuFLO}                     & 246\%               & 137\%            & 9.2\%        & 12.6\%  	& 11.7\%       & 10.4\%       & 13.5\%        & 13.1\%       \\
\textbf{Tamaraw}                   & 328\%               & 242\%            & 11.8\%       & 11.8\%		& 12.9\%       & 9.7\%        & 16.8\%        & 11.0\%       \\
\textbf{WTF-PAD}                   & 64\%                & 0\%              & 36.9\%       & \textbf{90.7\%} 	& 60.8\%  & 16.0\%       & 60.3\%        & 69.0\%       \\
\textbf{Walkie-Talkie}             & 31\%                & 34\%             & 23.1\%       & 49.7\%     & 45.8\%     & 20.2\%       & 38.4\%        & 7.0\%       \\ \hline
\end{tabular}

\end{table*}

\subsection{Training Cost}
We now examine the training time for WF attacks using DL in comparison to state-of-the-art WF attacks. We found that with GPU acceleration by using NVIDIA GTX 1070 with 8 GB of GPU Memory, SDAE required 16 minutes for training (13 minutes for pre-training and 3 minutes for fine-tuning processes), DF required 64 minutes for 30-epoch training. The relatively simpler AWF model requires 4 minutes for 30-epoch training. Without a GPU, SDAE required 96 minutes, DF required approximately 10 hours, and AWF required 1 hour.  
For training the other attacks, we found it required 12.5 hours for $k$-NN, 57 hours for CUMUL (parallelized with 4 processes), and 1 hour for $k$-FP. 
Overall, SDAE, DF and AWF have reasonable training times, particularly when using a GPU.


\subsection{Closed-world Evaluation on the Defended Dataset}\label{closed-world-defended}
We next examine the performance of WF attacks against Tor traffic with defenses in the closed-world scenario. It is important to note that the attacker needs to train the classifiers with defended datasets to perform this attack. As mentioned in Section~\ref{sec:related}, several WF defenses have been proposed that they can reduce the accuracy of state-of-the-art WF attacks to less than 50\%. Notably, WTF-PAD and Walkie-Talkie offer both effective protection and reasonable overheads, such that they are realistic for adoption in Tor. With our larger dataset, we conduct an evaluation on SDAE, DF, AWF and prior attacks against these defenses, as well as BuFLO and Tamaraw.  

Table \ref{accuracy-close-world-def} shows the overheads of each defense and the accuracy of the attacks against defended datasets. BuFLO and Tamaraw, the two high-overhead defenses, hold up well with less than 17\% accuracy. The attacks also manage at most 49.70\% accuracy against Walkie-Talkie due to symmetric collisions. A surprising result is that DF achieves over 90\% accuracy against WTF-PAD. Our tests of WTF-PAD showed 64\% overhead, which means that there was more padding on average than in the Juarez et al.'s study~\cite{JR2016}, and yet the attack was successful. More generally, it seems that the larger amount of traces per site compared to the original WTF-PAD evaluation has played a role in the higher accuracy attained by the attack. For example, $k$-FP achieved nearly 69\% accuracy in our experiment, whereas Hayes and Danezis tested $k$-FP against their own implementation of adaptive padding and obtained 30\% accuracy~\cite{Hayes2016}.

DF significantly outperforms AWF on the dataset protected by WTF-PAD, with a much larger gap in performance than observed on the undefended dataset. We believe that the deeper network is able to better extract useful features in the WTF-PAD data that the AWF model is unable to find, leading to this result. The model architecture in DF plays a key role in its flexibility to generalize to defended traffic.

We note that the overheads for BuFLO and Tamaraw are higher than reported in prior work at 246\% and 328\% bandwidth overheads, respectively. Furthermore, we found that the larger the dataset, the greater the packet timing variance is, which is fundamental to determine the padding rate. Also, Tamaraw has higher overheads than BuFLO, which contradicts the purposed intended with its design and the overheads reported in previous evaluations. The cause of this is a greater amount of padding after the transmission has finished in Tamaraw compared to BuFLO. BuFLO stops padding immediately after the transmission has finished, as long as the transmission has lasted for longer than ten seconds, which is the case for most of the traces in our dataset.

With such heavy overheads, BuFLO and Tamaraw are not practical to deploy as a WF defense in Tor. WTF-PAD and Walkie-Talkie have lower overheads, and Tor Project developers have already shown an interest in deploying {\em adaptive padding} as a possible defense~\cite{Perry2013,Perry2015}. We thus select WTF-PAD and Walkie-Talkie for our open-world evaluation.  
 
\subsection{Open-world Evaluation}
\label{OW-Evaluation}
We now evaluate the performance of the attack in the more realistic open-world setting.
As mentioned in Section~\ref{threatmodel}, in the open-world scenario, the adversary not only classifies traffic traces based on a limited set of monitored sites, but he must also distinguish whether the trace comes from a monitored site or an unmonitored one.

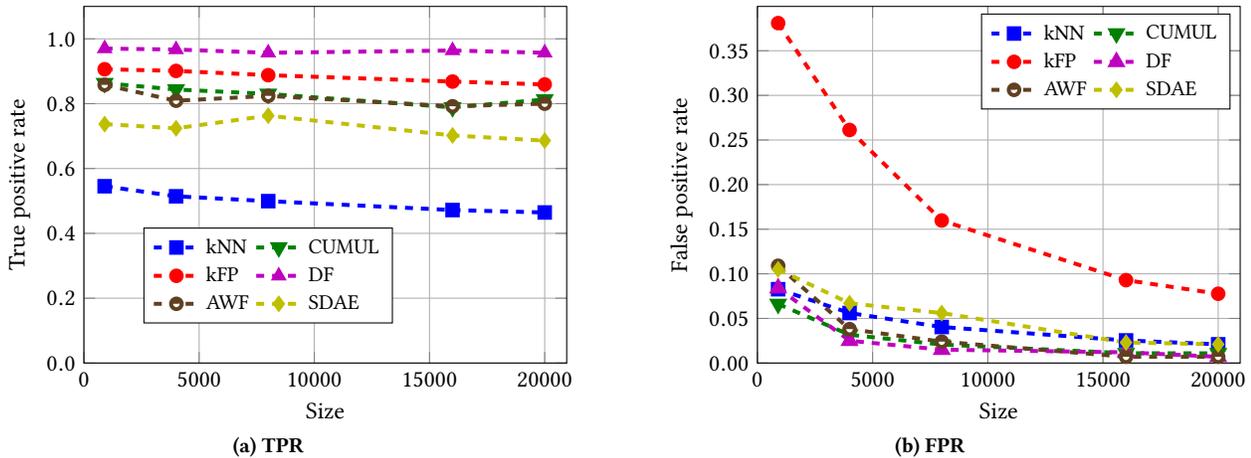
\begin{figure*}[th]
        \centering
        \hspace*{-1.8cm}
        \begin{subfigure}[ht]{.4\textwidth}
%
%
%
\begin{tikzpicture}

\definecolor{color1}{rgb}{0.75,0.75,0}
\definecolor{color0}{rgb}{0.75,0,0.75}

\begin{axis}[
xlabel={Size},
ylabel={True positive rate},
xmin=0, xmax=20955,
ymin=0, ymax=1.1,
width=\largefigurewidth,
height=180,
xtick={0,5000,10000,15000,20000,25000},
xticklabels={$0$,$5000$,$10000$,$15000$,$20000$,},
ytick={0,0.2,0.4,0.6,0.8,1},
yticklabels={$0.0$,$0.2$,$0.4$,$0.6$,$0.8$,$1.0$},
scaled ticks=false,
xmajorgrids,
x grid style={white!69.01960784313725!black},
ymajorgrids,
y grid style={white!69.01960784313725!black},
legend columns=2,
legend style={at={(0.64,0.38),font=\small}},
legend entries={{kNN},{CUMUL},{kFP},{DF},{AWF},{SDAE}},
legend cell align={left}
]
\addplot [thick, blue, dashed, mark=square*, mark size=2, mark options={solid},line width=\LineWidth]
table {%
900 0.545368421052632
4000 0.514315789473684
8000 0.499368421052632
16000 0.471894736842105
20000 0.464421052631579
};
\addplot [thick, green!50.0!black, dashed, mark=triangle*, mark size=2.5, mark options={solid,rotate=180},line width=\LineWidth]
table {%
900 0.863157894736842
4000 0.843684210526316
8000 0.830105263157895
16000 0.788631578947368
20000 0.812736842105263
};
\addplot [thick, red, dashed, mark=otimes*, mark size=2, mark options={solid},line width=\LineWidth]
table {%
900 0.906210526315789
4000 0.901052631578947
8000 0.887894736842105
16000 0.868210526315789
20000 0.859368421052632
};
\addplot [thick, color0, dashed, mark=triangle*, mark size=2, mark options={solid},line width=\LineWidth]
table {%
900 0.97
4000 0.967
8000 0.957
16000 0.964
20000 0.957
};
\addplot [thick, brown!50.0!black, dashed, mark=halfcircle*, mark size=2, mark options={solid},line width=\LineWidth]
table {%
900	0.858
4000 0.809
8000 0.824
16000 0.792
20000 0.8

};
\addplot [thick, color1, dashed, mark=diamond*, mark size=2, mark options={solid},line width=\LineWidth]
table {%
900 0.737
4000 0.724
8000 0.763
16000 0.702
20000 0.686
};
\end{axis}

\end{tikzpicture}
          \vspace*{-5mm}
          \caption{TPR}
          \label{fig:TPR_OW_NoDef}
        \end{subfigure}\hspace*{1.6cm}
        \begin{subfigure}[ht]{.4\textwidth}
%
%
%
\begin{tikzpicture}

\definecolor{color1}{rgb}{0.75,0.75,0}
\definecolor{color0}{rgb}{0.75,0,0.75}

\begin{axis}[
xlabel={Size},
ylabel={False positive rate},
xmin=0, xmax=20955,
ymin=0, ymax=0.3997,
width=\largefigurewidth,
height=180,
xtick={0,5000,10000,15000,20000,25000},
xticklabels={$0$,$5000$,$10000$,$15000$,$20000$,},
ytick={0,0.05,0.1,0.15,0.2,0.25,0.3,0.35,0.4},
yticklabels={$0.00$,$0.05$,$0.10$,$0.15$,$0.20$,$0.25$,$0.30$,$0.35$,$0.40$},
scaled ticks=false,
xmajorgrids,
x grid style={white!69.01960784313725!black},
ymajorgrids,
legend columns=2,
y grid style={white!69.01960784313725!black},
legend style={at={(0.98,0.98),font=\small}},
legend entries={{kNN},{CUMUL},{kFP},{DF},{AWF},{SDAE}},
legend cell align={left}
]
\addplot [thick, blue, dashed, mark=square*, mark size=2, mark options={solid},line width=\LineWidth]
table {%
900 0.08275
4000 0.05605
8000 0.0404
16000 0.02535
20000 0.021
};
\addplot [thick, green!50.0!black, dashed, mark=triangle*, mark size=2.5, mark options={solid,rotate=180},line width=\LineWidth]
table {%
900 0.06615
4000 0.0317
8000 0.0207
16000 0.011
20000 0.0109
};
\addplot [thick, red, dashed, mark=otimes*, mark size=2, mark options={solid},line width=\LineWidth]
table {%
900 0.381
4000 0.2612
8000 0.15975
16000 0.0929
20000 0.07765
};
\addplot [thick, color0, dashed, mark=triangle*, mark size=2.5, mark options={solid},line width=\LineWidth]
table {%
900 0.084
4000 0.025
8000 0.015
16000 0.012
20000 0.007
};
\addplot [thick, brown!50.0!black, dashed, mark=halfcircle*, mark size=2, mark options={solid},line width=\LineWidth]
table {%
900	0.109
4000 0.038
8000 0.024
16000 0.007
20000 0.007

};
\addplot [thick, color1, dashed, mark=diamond*, mark size=2, mark options={solid},line width=\LineWidth]
table {%
900 0.105
4000 0.067
8000 0.056
16000 0.023
20000 0.021
};
\end{axis}

\end{tikzpicture}
          \vspace*{-5mm}
          \caption{FPR} 
          \label{fig:FPR_OW_NoDef}
        \end{subfigure}\hspace{0.1cm}
        \caption{{\bf Open World:} The impact of the amount of unmonitored training data on TPR and FPR (Non-defended dataset).}\label{fig:tp-fp-size}
\end{figure*}



In our evaluation, we assess the performance of classifiers in the open-world scenario on each model by showing 
true positive rate (TPR) and false positive rate (FPR), but also with
precision and recall curves, recommended in the WF literature~\cite{JR2016,Panchenko2016} as more appropriate metrics for the open-world evaluation than TPR and FPR. The size of the monitored and unmonitored sets are heavily unbalanced, so using only TPR and FPR can lead to incorrect interpretation due to the base-rate fallacy. 
We also provide ROC curves in Appendix~\ref{appendix:OWResult}.

\paragraphX{Standard Model.}
In previous studies on WF in the open-world setting~\cite{Wang2014,Hayes2016,Panchenko2016}, it has been assumed that if the attacker included unmonitored traces when training the classifier, it could help the classifier better distinguish between monitored and unmonitored traces. This assumption is also common in machine learning, and we thus call it the Standard model.
Fundamentally, the process of training and creating datasets used during open-world evaluation is the same as in the closed-world scenario, except that we additionally train on unmonitored websites traces as another class. To investigate the impact of more training data on the performance of the classifiers in the open-world scenario, we train the classifier with different portions of the unmonitored dataset.

In our open-world evaluation, we use the prediction probability to classify the input traces. In particular, if the input trace is a monitored website trace and the maximum output probability belongs to any monitored site and is greater than a threshold, we consider this as a {\em true positive}. We used different thresholds for different WF attacks. We selected the thresholds for each WF attack such that they have high TPR and low FPR. Figure~\ref{fig:open-world-roc} in Appendix~\ref{appendix:OWResult} shows examples of ROC curves for WF attacks against Non-defended, WTF-PAD, and W-T datasets. Following the experimental procedures of Rimmer et al.~\cite{Rimmer2018} and Panchenko et al.~\cite{Panchenko2016},  we focus on the binary results of whether the input trace is classified as monitored (predicted to be in any of the monitored classes) or unmonitored. Note that if the trace is determined to be monitored, the attacker can then use the multi-class classification to predict which website the user has actually visited.

$k$-NN and $k$-FP attacks use the {\em k-nearest neighbors} algorithm in their predictions and do not output the probability of predictions. For these attacks, we consider the prediction probability of a site as the fraction of the nearest neighbors belonging to that site among the $k$ nearest neighbors. We explored the performance of these attacks as the value of $k$ varies from 2 to 10. We found that above $k=5$, the TPR and FPR do not change significantly. For our open-world evaluation, we used $k=6$ in both $k$-NN and $k$-FP.  

\subsubsection{Results}

We first evaluate efficacy of our WF attack in the Standard model as amounts of unmonitored training data varies and compare it with other state-of-the-art WF attacks on the non-defended traces. Our training set in this experiment contains 85,500 monitored traces (900 instances for each of 95 monitored sites) and we varied the number of unmonitored sites from 900 to 20,000 sites (one instance for each). Our testing set includes 9500 monitored traces (100 instances for 95 monitored sites) and 20,000 unmonitored traces (one instance for 20,000 unmonitored sites). Note that the 20,000 unmonitored sites in the testing are different from those in the training.

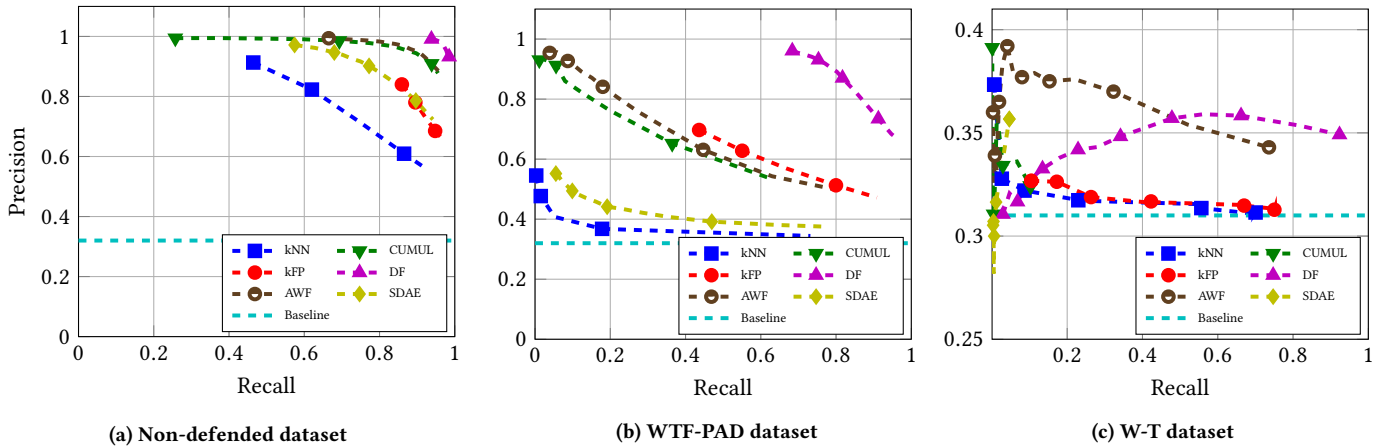
\begin{figure*}[t]
        \centering
        \hspace*{-.8cm}
        \begin{subfigure}[ht]{.34\textwidth}
%
%
%
\begin{tikzpicture}

\definecolor{color1}{rgb}{0.75,0.75,0}
\definecolor{color0}{rgb}{0.75,0,0.75}
\definecolor{color2}{rgb}{0,0.75,0.75}

\begin{axis}[
xlabel={Recall},
ylabel={Precision},
xmin=0, xmax=1,
ymin=0, ymax=1.1,
width=\smallfigurewidth,
height=170,
xtick={0,0.2,0.4,0.6,0.8,1},
ytick={0,0.2,0.4,0.6,0.8,1,1.2},
xmajorgrids,
x grid style={white!69.01960784313725!black},
ymajorgrids,
legend columns=2,
legend style={at={(0.98,0.31),font=\tiny}},
y grid style={white!69.01960784313725!black},
legend entries={{kNN},{CUMUL},{kFP},{DF},{AWF},{SDAE},{Baseline}},
legend cell align={left}
]
\addplot [thick, blue, dashed, mark=square*, mark size=\MarkerSize, mark options={solid},line width=\LineWidth,mark repeat={4}]
table {%
0.464421052631579 0.913079470198676
0.464421052631579 0.913079470198676
0.464421052631579 0.913079470198676
0.464421052631579 0.913079470198676
0.620210526315789 0.822905027932961
0.620210526315789 0.822905027932961
0.746421052631579 0.715612069835503
0.746421052631579 0.715612069835503
0.865263157894737 0.609340252038547
0.917894736842105 0.564474365613672
};
\addplot [thick, green!50.0!black, dashed, mark=triangle*, mark size=\MarkerSize, mark options={solid,rotate=180},line width=\LineWidth,mark repeat={4}]
table {%
0.257263157894737 0.994304312449146
0.36778947368421 0.994591517221748
0.483157894736842 0.992861778066191
0.597473684210526 0.990575916230366
0.693157894736842 0.985483388207124
0.773368421052632 0.977904964727805
0.837684210526316 0.967067687446834
0.897578947368421 0.943983172810805
0.938526315789474 0.909146527990211
0.955473684210526 0.877598375713043
};
\addplot [thick, red, dashed, mark=otimes*, mark size=\MarkerSize, mark options={solid},line width=\LineWidth,mark repeat={4}]
table {%
0.859368421052632 0.84017700936503
0.859368421052632 0.84017700936503
0.859368421052632 0.84017700936503
0.859368421052632 0.84017700936503
0.895263157894737 0.780346820809249
0.895263157894737 0.780346820809249
0.922210526315789 0.729596935376416
0.922210526315789 0.729596935376416
0.947789473684211 0.685235920852359
0.961578947368421 0.666545056548705
};
\addplot [thick, color0, dashed, mark=triangle*, mark size=\MarkerSize, mark options={solid},line width=\LineWidth,mark repeat={15}]
table {%
0.937894736842105 0.991101223581758
0.942105263157895 0.990592141671278
0.946315789473684 0.98910771261965
0.950947368421053 0.98743032025358
0.954315789473684 0.985649054142205
0.957368421052632 0.98398788272206
0.959789473684211 0.982966796032773
0.962315789473684 0.979954979097438
0.965894736842105 0.976481855911461
0.968315789473684 0.972718621127207
0.970842105263158 0.968497322272393
0.973157894736842 0.962820245782129
0.977157894736842 0.953765539915751
0.980421052631579 0.941283476503284
0.985052631578947 0.932164558222931
0.985052631578947 0.932164558222931
0.985052631578947 0.932164558222931
0.985052631578947 0.932164558222931
0.985052631578947 0.932164558222931
0.985052631578947 0.932164558222931
0.985052631578947 0.932164558222931
0.985052631578947 0.932164558222931
0.985052631578947 0.932164558222931
};
\addplot [thick, brown!50.0!black, dashed, mark=halfcircle*, mark size=\MarkerSize, mark options={solid},line width=\LineWidth,mark repeat={15}]
table {%
0.665 0.994
0.691 0.992
0.721 0.99
0.749 0.988
0.775 0.986
0.8 0.983
0.82 0.979
0.841 0.975
0.863 0.969
0.882 0.958
0.905 0.948
0.922 0.929
0.944 0.901
0.962 0.878
0.968 0.872

};
\addplot [thick, color1, dashed, mark=diamond*, mark size=\MarkerSize, mark options={solid},line width=\LineWidth,mark repeat={4}]
table {%
0.574315789473684 0.971682991985752
0.601157894736842 0.96665538253216
0.629052631578947 0.961699388477631
0.654947368421053 0.954148136788836
0.67978947368421 0.94650447017441
0.701894736842105 0.938626126126126
0.725894736842105 0.926259234385494
0.749473684210526 0.916698854126432
0.772210526315789 0.901782421634911
0.796315789473684 0.88314265701611
0.82221052631579 0.861000881834215
0.855789473684211 0.830778663396689
0.896526315789474 0.786208806424813
0.931789473684211 0.736071844337269
0.941789473684211 0.723516092511726
};
\addplot [semithick, color2, dashed,line width=\LineWidth]
table {%
0 0.32
1 0.32
};
\end{axis}

\end{tikzpicture}
          \vspace{-1.0em}
          \caption{Non-defended dataset}          
          \label{fig:PRC_OW_NoDef}
        \end{subfigure}\hspace{0.37cm}
        \begin{subfigure}[ht]{.34\textwidth}
%
%
%
\begin{tikzpicture}

\definecolor{color1}{rgb}{0.75,0.75,0}
\definecolor{color0}{rgb}{0.75,0,0.75}
\definecolor{color2}{rgb}{0,0.75,0.75}

\begin{axis}[
xlabel={Recall},
xmin=0, xmax=1,
ymin=0, ymax=1.1,
width=\smallfigurewidth,
height=170,
xtick={0,0.2,0.4,0.6,0.8,1},
ytick={0,0.2,0.4,0.6,0.8,1},
xmajorgrids,
x grid style={white!69.01960784313725!black},
ymajorgrids,
y grid style={white!69.01960784313725!black},
legend columns=2,
legend style={at={(0.98,0.31),font=\tiny}},
legend entries={{kNN},{CUMUL},{kFP},{DF},{AWF},{SDAE},{Baseline}},
legend cell align={left}
]
\addplot [thick, blue, dashed, mark=square*, mark size=\MarkerSize, mark options={solid},line width=\LineWidth,mark repeat={4}]
table {%
0.00315789473684211 0.545454545454545
0.00315789473684211 0.545454545454545
0.00315789473684211 0.545454545454545
0.00315789473684211 0.545454545454545
0.0152631578947368 0.476973684210526
0.0152631578947368 0.476973684210526
0.0514736842105263 0.407839866555463
0.0514736842105263 0.407839866555463
0.178315789473684 0.36786102062975
0.731789473684211 0.344192494306367
};
\addplot [thick, green!50.0!black, dashed, mark=triangle*, mark size=\MarkerSize, mark options={solid,rotate=180},line width=\LineWidth,mark repeat={4}]
table {%
0.0112631578947368 0.930434782608696
0.0196842105263158 0.935
0.0268421052631579 0.930656934306569
0.0386315789473684 0.924433249370277
0.0558947368421053 0.912371134020619
0.0796842105263158 0.860227272727273
0.123473684210526 0.82258064516129
0.198631578947368 0.762424242424242
0.364526315789474 0.652165725047081
0.627157894736842 0.534877457581471
};
\addplot [thick, red, dashed, mark=otimes*, mark size=\MarkerSize, mark options={solid},line width=\LineWidth,mark repeat={4}]
table {%
0.436105263157895 0.697122665320545
0.436105263157895 0.697122665320545
0.436105263157895 0.697122665320545
0.436105263157895 0.697122665320545
0.550526315789474 0.628001921229587
0.550526315789474 0.628001921229587
0.668526315789474 0.568882121103547
0.668526315789474 0.568882121103547
0.79978947368421 0.512858589267634
0.908526315789474 0.472052067381317
};
\addplot [thick, color0, dashed, mark=triangle*, mark size=\MarkerSize, mark options={solid},line width=\LineWidth,mark repeat={4}]
table {%
0.683684210526316 0.960798816568047
0.699789473684211 0.953938872148084
0.718736842105263 0.947675225537821
0.736 0.940796555435953
0.753263157894737 0.930680192482768
0.770526315789474 0.918905347727843
0.785473684210526 0.905582524271845
0.802 0.891319606925597
0.817157894736842 0.870096390943735
0.834631578947368 0.846753524134985
0.854631578947368 0.819935366592608
0.878210526315789 0.785371364021463
0.912421052631579 0.734202947653735
0.946947368421053 0.683950429559796
0.964105263157895 0.666788002329645
};
\addplot [thick, brown!50.0!black, dashed, mark=halfcircle*, mark size=\MarkerSize, mark options={solid},line width=\LineWidth,mark repeat={4}]
table {%
0.039 0.954
0.047 0.957
0.059 0.956
0.072 0.948
0.087 0.927
0.104 0.912
0.123 0.892
0.151 0.871
0.18 0.841
0.221 0.804
0.274 0.758
0.343 0.707
0.447 0.631
0.63 0.543
0.816 0.494

};
\addplot [thick, color1, dashed, mark=diamond*, mark size=\MarkerSize, mark options={solid},line width=\LineWidth,mark repeat={4}]
table {%
0.0553684210526316 0.551941238195173
0.0637894736842105 0.540588760035682
0.0736842105263158 0.523952095808383
0.0863157894736842 0.509633312616532
0.0990526315789474 0.49422268907563
0.115368421052632 0.482182138143423
0.135789473684211 0.470459518599562
0.161473684210526 0.458594917787743
0.191684210526316 0.441561590688652
0.234 0.43384074941452
0.285789473684211 0.421387552382431
0.357789473684211 0.407114624505929
0.469789473684211 0.391731765118933
0.646 0.380659967745937
0.770315789473684 0.37516661540039
};
\addplot [semithick, color2, dashed,line width=\LineWidth]
table {%
0 0.32
1 0.32
};
\end{axis}

\end{tikzpicture}
          \caption{WTF-PAD dataset}
          \label{fig:PRC_OW_WTF}          
        \end{subfigure}\hspace{-0.2cm}        
        \begin{subfigure}[ht]{.34\textwidth}
%
%
%
\begin{tikzpicture}

\definecolor{color1}{rgb}{0.75,0.75,0}
\definecolor{color0}{rgb}{0.75,0,0.75}
\definecolor{color2}{rgb}{0,0.75,0.75}

\begin{axis}[
xlabel={Recall},
xmin=0.0005, xmax=1,
ymin=0.25,ymax = 0.41,
width=\smallfigurewidth,
height=170,
xmajorgrids,
x grid style={white!69.01960784313725!black},
ymajorgrids,
legend columns=2,
y grid style={white!69.01960784313725!black},
legend style={at={(0.98,0.31),font=\tiny}},
legend entries={{kNN},{CUMUL},{kFP},{DF},{AWF},{SDAE},{Baseline}},
legend cell align={left}
]
\addplot [thick, blue, dashed, mark=square*, mark size=\MarkerSize, mark options={solid},line width=\LineWidth,mark repeat={4}]
table {%
0.00622222222222222 0.373333333333333
0.00622222222222222 0.373333333333333
0.00622222222222222 0.373333333333333
0.00622222222222222 0.373333333333333
0.00622222222222222 0.373333333333333
0.00622222222222222 0.373333333333333
0.00622222222222222 0.373333333333333
0.0263333333333333 0.327800829875519
0.0263333333333333 0.327800829875519
0.0263333333333333 0.327800829875519
0.0263333333333333 0.327800829875519
0.0263333333333333 0.327800829875519
0.0866666666666667 0.32204789430223
0.0866666666666667 0.32204789430223
0.0866666666666667 0.32204789430223
0.229555555555556 0.317455439459127
0.229555555555556 0.317357910906298
0.229222222222222 0.317189421894219
0.561666666666667 0.315405253634492
0.561666666666667 0.314757160647572
0.556333333333333 0.313564629258517
0.699444444444444 0.310741435482279
0.705888888888889 0.312586105097422
0.700777777777778 0.311641466548078
0.700444444444444 0.311416292051573
0.705777777777778 0.312783139649399
0.700333333333333 0.311044216344256
};
\addplot [thick, green!50.0!black, dashed, mark=triangle*, mark size=\MarkerSize, mark options={solid,rotate=180},line width=\LineWidth,mark repeat={4}]
table {%
0.001 0.391304347826087
0.001 0.391304347826087
0.002 0.321428571428571
0.00288888888888889 0.333333333333333
0.00455555555555556 0.310606060606061
0.005 0.306122448979592
0.0128888888888889 0.355828220858896
0.0137777777777778 0.358381502890173
0.0292222222222222 0.33418043202033
0.0538888888888889 0.335640138408304
0.0664444444444444 0.336143901068016
0.0997777777777778 0.325009048136084
0.101777777777778 0.324362606232295
0.102222222222222 0.324057766819303
0.102777777777778 0.324220119172801
0.102777777777778 0.324106517168886
0.102777777777778 0.324106517168886
};
\addplot [thick, red, dashed, mark=otimes*, mark size=\MarkerSize, mark options={solid},line width=\LineWidth,mark repeat={4}]
table {%
0.104666666666667 0.326742976066597
0.104666666666667 0.326742976066597
0.104666666666667 0.326742976066597
0.104666666666667 0.326742976066597
0.104666666666667 0.326742976066597
0.104666666666667 0.326742976066597
0.104666666666667 0.326742976066597
0.172333333333333 0.326320218809173
0.172333333333333 0.326320218809173
0.172333333333333 0.326320218809173
0.172333333333333 0.326320218809173
0.172333333333333 0.326320218809173
0.263333333333333 0.318891280947255
0.263333333333333 0.318891280947255
0.263333333333333 0.318891280947255
0.422333333333333 0.316144057223655
0.422777777777778 0.316792939805179
0.422555555555556 0.316468336523259
0.670444444444444 0.315042029969195
0.671 0.31523725009135
0.67 0.31476744793026
0.750666666666667 0.313212795549374
0.749444444444444 0.31324014303627
0.750333333333333 0.312740239892558
0.750444444444444 0.312772066314717
0.752777777777778 0.314166473452353
0.752888888888889 0.313805399898115
};
\addplot [thick, color0, dashed, mark=triangle*, mark size=\MarkerSize, mark options={solid},line width=\LineWidth,mark repeat={2}]
table {%
0.0293333333333333 0.310588235294118
0.0468888888888889 0.320668693009119
0.068 0.316606311433006
0.0975555555555556 0.326758466691478
0.134444444444444 0.332600329851567
0.178444444444444 0.337820782498948
0.228888888888889 0.341851974775971
0.279888888888889 0.343141261408527
0.341555555555556 0.348249688455874
0.405777777777778 0.352373600926283
0.478222222222222 0.357059897129584
0.561333333333333 0.359036315826878
0.661888888888889 0.358337343599615
0.814666666666667 0.353605015673981
0.923444444444444 0.349143001176273
};
\addplot [thick, brown!50.0!black, dashed, mark=halfcircle*, mark size=\MarkerSize, mark options={solid},line width=\LineWidth,mark repeat={2}]
table {%
0.003	0.36
0.005	0.326
0.008	0.339
0.012	0.332
0.019	0.365
0.028	0.383
0.041	0.392
0.057	0.379
0.08	0.377
0.113	0.379
0.153	0.375
0.22	0.376
0.323	0.37
0.52	0.354
0.736	0.343

};
\addplot [thick, color1, dashed, mark=diamond*, mark size=\MarkerSize, mark options={solid},line width=\LineWidth,mark repeat={4}]
table {%
0.00322222222222222 0.305263157894737
0.00333333333333333 0.303030303030303
0.00366666666666667 0.308411214953271
0.00411111111111111 0.310924369747899
0.00433333333333333 0.307086614173228
0.00466666666666667 0.295774647887324
0.00466666666666667 0.281879194630872
0.00522222222222222 0.286585365853659
0.006 0.3
0.00655555555555556 0.307291666666667
0.00755555555555556 0.306306306306306
0.00877777777777778 0.311023622047244
0.0111111111111111 0.316455696202532
0.0156666666666667 0.321917808219178
0.0463333333333333 0.356715141146279
0.0463333333333333 0.356715141146279
0.0463333333333333 0.356715141146279
0.0463333333333333 0.356715141146279
0.0463333333333333 0.356715141146279
0.0463333333333333 0.356715141146279
0.0463333333333333 0.356715141146279

};
\addplot [semithick, color2, dashed,line width=\LineWidth]
table {%
0.0005 0.31
1 0.31
};
\end{axis}

\end{tikzpicture}
           \caption{W-T dataset}
          \label{fig:PRC_OW_WT}
        \end{subfigure} 
        \vspace{-0.3em}
         \caption{{\bf Open World:} Precision-Recall curves.}\label{fig:open-world-prc}
\end{figure*}

As shown in Figures~\ref{fig:TPR_OW_NoDef} and \ref{fig:FPR_OW_NoDef}, the TPR tends to slightly decrease with the reduction of FPR as the size of unmonitored training data increase for all the WF attacks. The results show that the DF model consistently performs best on both TPR and FPR, with 0.957 TPR and 0.007 FPR for 20,000 unmonitored training sites. $k$-NN has the lowest TPR and $k$-FP has the highest FPR. The DF, CUMUL and AWF have the same FPR trend as the training size increases, but DF has higher TPR than CUMUL and AWF over all the training sizes. 

Our results show that as we increase the size of the unmonitored class in the training, the FPR drops and it reaches its lowest amount at size 20,000. In the next experiment, we fix the number of training samples for the unmonitored class to 20,000 and we evaluate the diagnostic ability of WF attacks as the discrimination threshold is varied. We next perform the experiment on our non-defended, WTF-PAD and Walkie-Talkie (W-T) datasets. As mentioned in Section~\ref{sec:data_collection}, for W-T, we cannot use the same dataset to W-T traces as it required to be directly captured from half-duplex connections from Tor browser. Our training set for the W-T evaluation contains 91,000 monitored traces (910 instances for each of 100 monitored sites) and we varied the number of unmonitored sites from 900 to 20,000 sites (one instance for each). Our testing set includes 9,000 traces (90 instances for each of 100 monitored sites) and 20,000 unmonitored traces (one instance for 20,000 unmonitored sites). In the following open-world experiments, we mainly focus on the precision and recall to avoid the base-rate fallacy as mentioned above. 

Figure~\ref{fig:open-world-prc} shows the precision-recall curves for WF attacks in our non-defended, WTF-PAD and W-T datasets. Precision-recall curves are used to represent the performance of the classifier as an alternative to ROC curves in imbalanced datasets. Imbalanced datasets have an impact on precision, an important metric to measure performance, however, ROC curves do not take precision into account. This choice is specially relevant in the open-world evaluation, as the size of the monitored set is typically orders of magnitude smaller than the unmonitored set and as such it should be represented in our testing set, thus leading to an imbalance~\cite{JR2016}.

As we see in the figure, the DF attack outperforms the other state-of-the-art WF attacks in all three cases. 
In the non-defended dataset, it is highly effective for any threshold.
The CUMUL and AWF attacks in Figure~\ref{fig:PRC_OW_NoDef} have high precision but a very wide range of recall, which means the attacks 
miss many monitored visits. For traffic defended by WTF-PAD, Figure~\ref{fig:PRC_OW_WTF} shows a reduction of both precision and recall for all WF attacks. The DF attacker does the best. Tuned for high precision, it achieves precision of 0.96 and recall of 0.68. Tuned for high recall, it reaches 0.67 precision and 0.96 recall. All the other WF attacks get close to the baseline (random guessing) as the threshold decreases. The result shows that the otherwise robust WTF-PAD is significantly undermined by the DF attack.

Figure~\ref{fig:PRC_OW_WT} shows the precision-recall curves for the W-T dataset. The attacks all perform quite poorly, with all except the DF attack close to the baseline. The DF attack does moderately better but still has a precision of less than 0.36 in all cases.

\subsection{A Deeper Look at W-T}
\paragraphX{Top-N prediction for closed-world W-T}
Wang and Goldberg explain that any attack using the main features of the state-of-the-art attacks can get at most 50\% accuracy against W-T~\cite{Wang2017}. In our closed-world results, the DF attack nearly reached this theoretical maximum. We now examine prediction probability for DF against W-T. We consider top-$N$ prediction, in which we look at not only the highest probability (Top-1 prediction), but also the top $N$ probability values. Surprisingly, we only need to look at the case of $N=2$. Top-2 prediction accuracy reaches 98.44\% accuracy. This likely means that DF is correctly selecting the real site and the decoy site and, as expected, having to guess randomly between them. We discuss the importance of this result in Section~\ref{sec:discussion}.

\paragraphX{Asymmetric Collision (Closed-World)}
W-T requires that the client create symmetric collisions between pairs of websites (site \textit{A} is molded with site \textit{B} and vice versa). Since this requires storing all the pairs, a simpler implementation would ignore this requirement and have the client random select the decoy site for each access, resulting in asymmetric collisions. In this setting, the DF attack is much more accurate at 87.2\%, compared to 49.7\% with symmetric collisions. This shows the importance of creating symmetric collisions in W-T.

\paragraphX{Asymmetric Collision (Open-World)}
We next evaluate the scenario that 10\% of the users do not follow the W-T guidelines, in that they visit a non-sensitive site and choose a non-sensitive site as a decoy instead of a sensitive one, and when they visit a sensitive site they choose a sensitive site as a decoy instead of a non-sensitive one. In this scenario, TPR is increased to 0.85 TPR, and FPR is significantly reduced to 0.23 TPR, compared to the case that all users strictly follow the procedure to create symmetric collisions which has 0.80 TPR and 0.76 FPR. Thus, the major goal of W-T to create the confusion between sensitive websites and non-sensitive websites could be undermined in some  scenarios. 
\section{Discussion}\label{sec:discussion}


The results of our study show that deep learning, and the DF approach in particular, is a powerful tool for WF attacks against Tor. Further, our results against defended traffic show that WTF-PAD is potentially vulnerable against deep-learning-based WF, with high precision even in the open-world setting.
Based on what we have observed during experimental evaluations, we now discuss several new directions in both improving the attacks and exploring designs for more effective defenses. 

\paragraphX{Improving Open-World Classification.}
In our study, we observed that designing the CNN architecture and tuning hyperarameters are specific to both the environment and input data. For example, the gap in performance between DF and AWF was much larger for the open-world setting than the closed world.
Additional exploration of models in the open-world scenario, such as the depth and number of convolutional layers, different filter sizes, or different dropout parameters, may yield improved results beyond what we have found so far. More training data may also help the classifier better distinguish between monitored and unmonitored pages. Our simple data format might be extended to include, for example, statistical timing information that is currently excluded.

Finally, we note that the attacker can perform a targeted attack on users in a {\em semi-open-world} setting, in which the targeted users can be profiled as likely going to a subset of sites. For example, if the user is known to only read one or two languages, then many sites in other languages can be eliminated from the set. Alternatively, a user's profile can help the attacker identify some likely sites for her interests, such that the classification of statistically similar monitored sites may be dismissed as likely false positives.


\paragraphX{Attack Costs.}
The time and effort needed to collect and train on large data sets can have practical implications for weaker attackers. Collecting large data sets as used in our experiments requires multiple PCs running for several days. Both Juarez et al.~\cite{Juarez2014} and Wang and Goldberg~\cite{Wang2016} show that after 10-14 days, the accuracy of WF attacks goes down significantly. A weak attacker might need to choose between limiting the scope of monitored sites, living with the results of using stale and inaccurate data, or using fewer training samples per site. We note that, even though deep learning works best with more data, DF performs well in the closed-world setting even with smaller datasets. Additionally, we found that while $k$-FP, AWF, and DF can be trained quickly on large datasets, $k$-NN and CUMUL do not scale well to larger data. In particular, due to hyperparameters grid search, CUMUL took us days to train. Further exploring the trade-offs between scalability and accuracy remain important areas for future research.

\paragraphX{WTF-PAD.}
As DF can break WTF-PAD with over 90\% accuracy in the closed-world setting, we now consider why the defense failed by examining the adaptive padding algorithm at the heart of WTF-PAD. Adaptive padding aims to detect large timing gaps between bursts and use padding to make these gaps less distinctive. While Juarez et al. showed that this is effective against prior WF attacks~\cite{JR2016}, DF can still detect patterns that remain after WTF-PAD is applied. When used in analyzing images, CNN can detect an object (e.g. a dog) anywhere in an image due to its use of convolutional layers with multiple filters. Similarly, DF can detect any small region or portion of distinguishing patterns, no matter where those patterns are located in the trace. Adaptive padding only randomly disrupts some patterns in the trace, leaving other patterns relatively unperturbed. 

\paragraphX{Walkie-Talkie.}
Walkie-Talkie (W-T) has an advantage over WTF-PAD, as it focuses directly on features used in WF attacks, and it seeks explicitly to create collisions. Indeed, W-T performed much better than WTF-PAD against DF, which would seem to make it a strong candidate for Tor. We note, however, that there are several downsides to deploying W-T that require further investigation to overcome:
\begin{smitemize}
\item It requires the directory server to collect and distribute to all clients a database of website patterns that can be used to set the padding patterns. The patterns need to be kept up to date to provide effective plausible deniability.
\item Half-duplex communication adds to the latency of fetching a site, 31\% according to Wang and Goldberg~\cite{Wang2017}, which is a direct cost to end-user performance in a system that is already slower than regular browsing.
\item According to Wang and Goldberg, the browser is expected to pair sensitive and non-sensitive pages and, ideally, pay attention to issues such as language to select realistic cover pages. To be most effective, then, the browser has to have a lot of context about the user and the nature of her activity, which is hard to build into the system.
\item Given that DF achieves very high Top-2 accuracy, the attacker can use auxiliary information such as language to guess the real site. Further, if the system does not assign a decoy site to a particular sensitive site or page (e.g. beyond the homepage of the site), then that site is effectively uncovered, because it will not be used as a decoy for any non-sensitive sites.
\end{smitemize}
\paragraphX{Alternative Defenses.} To improve effectiveness against DF without requiring extensive interaction with the browser, defenses could apply \textit{adversarial machine learning}~\cite{Goodfellow14, Carlini17} to generate the adversarial website traces to confuse the classifier. This is challenging to do compared to adversarial machine learning in image processing tasks, since the web trace is happening live, where the Tor client does not know the full picture of the trace in advance. Further, Tor is limited in how it can manipulate the trace---it can add dummy packets and delay packets but not delete packets or speed them up. Addressing these challenges would be interesting for future work. 

\section{Conclusion}\label{sec:conclusion}
In this study, we 
investigated the performance of WF using deep learning techniques in both the closed-world scenario and the more realistic open-world scenario. We proposed a WF attack called Deep Fingerprinting (DF) using a sophisticate design based on a CNN for extracting features and classification. Our closed-world results show that the DF attack outperforms other state-of-the-art WF attacks, including better than 90\% accuracy on traffic defended by WTF-PAD. We also performed open-world experiments, including the first open-world evaluation of WF attacks using deep learning against defended traffic. On undefended traffic, the DF attack attains a 0.99 precision and a 0.94 recall, while against WTF-PAD, it reaches a 0.96 precision and a 0.68 recall. Finally, we provided a discussion on our results along with suggestions for further investigation. 

Overall, our study reveals the need to improve WF defenses to be more robust against attacks using deep learning, as attacks only get better, and we have already identified several directions to improve the DF attack further.

\begin{acks}
We thank the anonymous reviewers for their helpful feedback. A special acknowledgement to Vera Rimmer for providing feedback that helped improve the paper. We appreciate the interesting discussions with Vera Rimmer, Dr.\ Leon Reznik and Igor Khokhlov that helped developing this paper.

This material is based upon work supported by the National Science Foundation under Grants No.\ CNS-1423163 and CNS-1722743. In addition, this work has been supported by the European Commission through KU Leuven BOF OT/13/070 and H2020-DS-2014-653497 PANORAMIX. Juarez is supported by a PhD fellowship of the Fund for Scientific Research - Flanders (FWO).

\end{acks}

\bibliographystyle{ACM-Reference-Format}
\bibliography{References}

\appendix
\begin{figure}[tbp]
  \centering
  \includegraphics[width=0.37\textwidth]{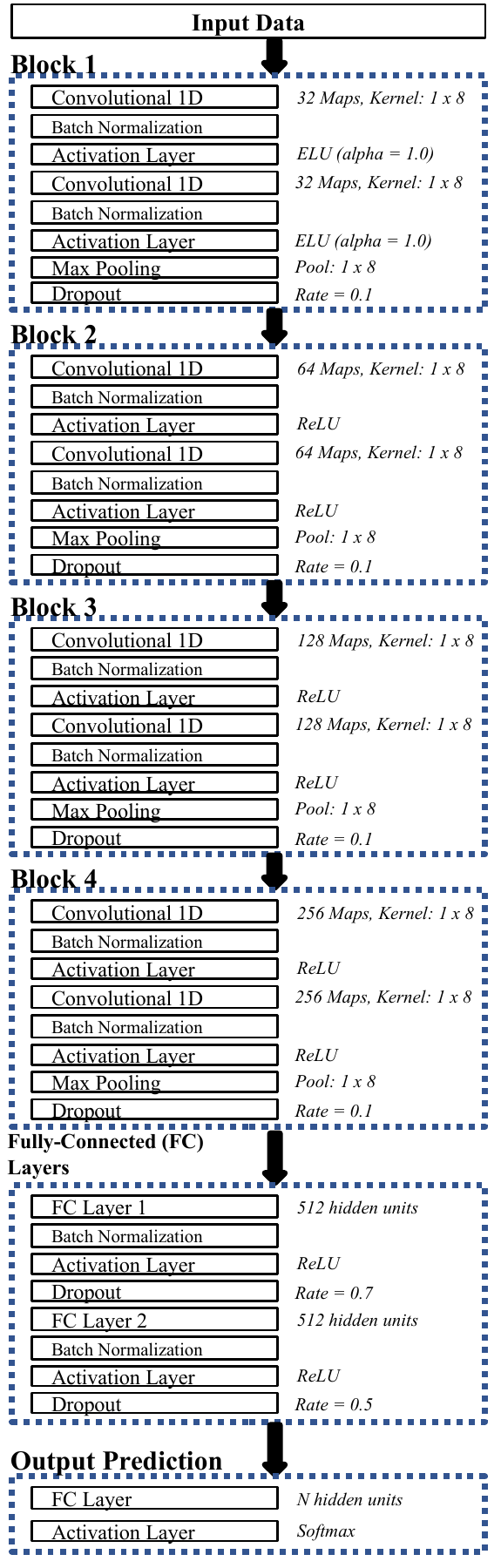}
  \caption{Our design of DF model's architecture used in WF attacks}
  \label{fig:CNN_full_model}  
\end{figure}

\section{Deep Fingerprinting (DF) Model's Architecture for WF attacks}\label{appendix:CNNModel}
One of the compelling properties for CNN is the transferability of the model. The transferability refers to the ability of the model to be used as a base model for similar tasks. Instead of training an entire CNN from scratch, the researcher can adapt the model to a similar task, specifically with a similar input format. 

In WF research, to the best of our knowledge, we are the first who provide the full technical details, guidelines and suggestions on how to implement our CNN-based DF model to perform WF attacks. In this section we provide details for our DF architecture and its hyperparameters to create our model, and to allow other researchers to apply it in their future work (see Figure~\ref{fig:CNN_full_model}): 

\paragraphX{Input Data.}
The input data for our DF model is the vector of packets' directions with length 5,000 (1 x 5,000). We initially tried adjusting the input dimension to be a matrix of shape similar to the matrices typically fed into CNNs for image recognition tasks (e.g., 50 x 100 pixels). The accuracy for 2D input was reasonably good, but slightly lower than 1D input. The major difference is training time: 1D input is significantly faster than 2D input, even though the total number of data points is the same for both input dimensions. We presume this difference results from tensor operations that have to deal with higher dimensions of data. We suggest that for the WF task, it is more appropriate to use 1D input as it is faster for training and provides better classification performance 

\paragraphX{Convolutional Layers (Block 1).}
Figure~\ref{fig:CNN_full_model} describes the architecture of our DF model divided by blocks, where a block comprises a set of convolutional layers, a batch normalization layer, a max pooling layer and a dropout layer. The first block in the DF is specially important due to its proximity to the input.

As we mentioned in Section~\ref{sec:Evaluation}, since the nature of our input is different to inputs considered in image recognition, we had to find an activation function that fits our input values. We chose the Exponential Linear Unit (ELU) because prior work has shown that it provides fast and accurate classification with negative inputs~\cite{Clevert2015,Mishkin2016}.
The results obtained from hyperparameters tuning suggested that applying ELU in the first two convolutional layers followed by ReLU with the rest of convolutional layers provides the best accuracy compared to only using ReLU. This suggests that ELU plays an important role in extracting hidden features from the input data.

\paragraphX{Dropout Regularization.}
CNN-based models are specially vulnerable to overfitting, an issue that might be easily overlooked by the developer. We applied a dropout technique to mitigate overfitting in the design of our DF model.
Our strategy to apply dropout was to embed in between feature extraction (Blocks 1-4) and classification (Fully-connected layers) using different rates. In feature extraction, we deployed dropout right after the max pooling layer in each block with 0.1 dropout rate. In addition, we added a dropout layer after each fully-connected layer with rate 0.7 and 0.5, respectively. As we observed from the hyperparameters tuning, the overfitting mostly arises at the fully-connected layers, and it is less problematic at the convolutional layers. Thus, we adjusted different dropout rates appropriately according to this observation.       

\paragraphX{Batch Normalization.}
We studied the technique to accelerate model learning called \textit{Batch Normalization (BN)}~\cite{Ioffe2015}. This technique provides benefits to improve the classification performance in order to, for instance, learn faster while maintaining or even increasing accuracy. Moreover, it also partially serves as a regulation method as well. Thus, we applied BN and dropout regularization together and obtained a boost in both performance and generalization. However, adding BN layers requires additional training time. We observed that it added around 100\% training time for each epoch compared to the model that did not apply BN. Yet, we believe it is worth applying BN, as the additional training time is compensated with a faster learning rate (it requires less number of epochs to reach the same level of accuracy) and can ultimately achieve higher testing accuracy. In our model, we applied BN right after every convolutional and fully-connected layers.

In conclusion, the researcher can apply this model and our suggestions to develop their own CNN-based model for WF. There are other details that we cannot describe here due to limit space including number of filters, kernel size, stride size and pool size. However, we will ensure that our implementation details, along with the source code and data used in this study, will be published on a website upon publication of this paper, so that researchers can reproduce our results.

\section{Attack Performance Metrics}\label{appendix:PerformanceMeasures}
In this section, we define and justify the metrics we have used to evaluate the success of the attacks. There are two scenarios under which WF attacks are evaluated: \textit{closed-world} and \textit{open-world}.

\balance
\subsection{Closed-world Evaluation}
In the closed-world scenario, we assume that the user is limited to visiting a fixed set of websites, and the attacker knows this set and can train his classifier on it. In this scenario, the success of the attacker is simply measured as the ratio of the number of correctly classified traces to the total number of traces, we call this ratio the attack's \textit{accuracy}.

\begin{equation} \label{accequ}
\begin{split}
Accuracy & = \frac{P_{correct}}{N} 
\end{split}
\end{equation}

\(P_{correct}\) is the total number of correct predictions. A correct prediction is defined as the output of the classifier matching the label of the website to which the test trace belongs. \({N}\) is the total number of instances in the test set. 

\subsection{Open-world Evaluation}
In the open-world scenario, the user may visit any of a large number of websites. Since the attacker cannot effectively train on so many sites, he selects a relatively small set to train his classifier on (the {\em monitored set}). For experimentation, we model the rest of the Web using a set of sites that the attacker does not try to identify with WF (the {\em unmonitored set}). Note that the unmonitored set is more than two orders of magnitude larger than the monitored set in our experiments.
As mentioned in Section~\ref{OW-Evaluation}, we measure \textit{Precision} and \textit{Recall} in this scenario.

\begin{equation} \label{Prec}
\begin{split}
Precision & = \frac{TP}{TP+FP} 
\end{split}
\end{equation}

\begin{equation} \label{Rec}
\begin{split}
Recall & = \frac{TP}{TP+FN} 
\end{split}
\end{equation}

\noindent Where:
\begin{smitemize}
\item \({TP}\) is the total number of test samples of monitored websites that are correctly classified as monitored websites.
\item \({TN}\) is the total number of test samples of unmonitored websites that are correctly classified as unmonitored websites.
\item \({FP}\) is the total number of test samples of unmonitored websites that are misclassified as monitored websites.
\item \({FN}\) is the total number of monitored websites that are misclassified as unmonitored websites.
\end{smitemize}

In addition, the attacker can measure \textit{precision} and \textit{recall} to tune the system. If his primary goal is to \emph{reliably} determine that a user has visited a particular monitored website, one can try to decrease false positives at the cost of true positives and thus increase the precision of the attack. On the other hand, if the attacker aims to cast a wide net and identify \emph{potential} visitors to the monitored web sites, then recall is more important, and the adversary should tune the system to increase true positives at the cost of additional false positives.

\section{Open-World ROC curve}\label{appendix:OWResult}
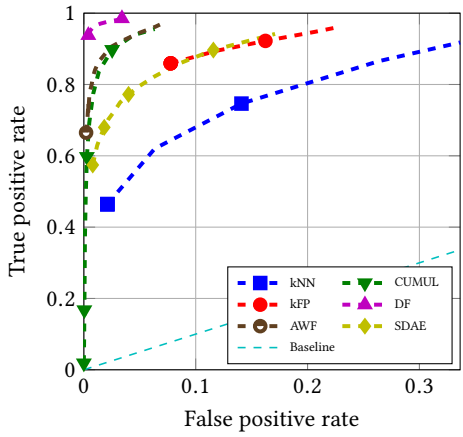
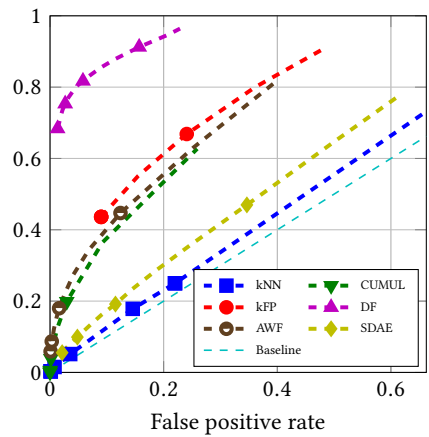
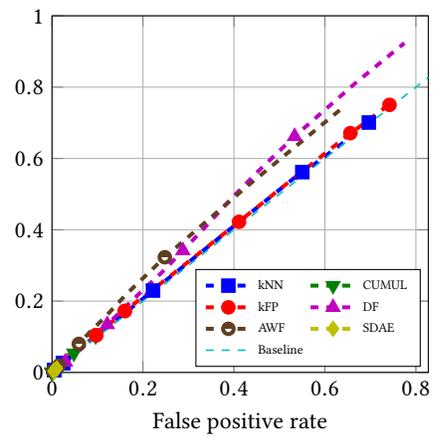
\begin{figure*}[t]
        \centering
        \hspace*{-.8cm}
        \begin{subfigure}[ht]{.34\textwidth}
%
%
%
\begin{tikzpicture}

\definecolor{color1}{rgb}{0.75,0.75,0}
\definecolor{color0}{rgb}{0.75,0,0.75}
\definecolor{color2}{rgb}{0,0.75,0.75}

\begin{axis}[
xlabel={False positive rate},
ylabel={True positive rate},
xmin=0, xmax=0.3364,
ymin=0, ymax=1,
width=\smallfigurewidth,
height=180,
xmajorgrids,
x grid style={white!69.01960784313725!black},
ymajorgrids,
y grid style={white!69.01960784313725!black},
legend entries={{kNN},{CUMUL},{kFP},{DF},{AWF},{SDAE},{Baseline}},
legend columns=2,
legend style={at={(0.98,0.29),font=\tiny}},
legend cell align={left}
]
\addplot [thick, blue, dashed, mark=square*, mark size=\MarkerSize,line width=\LineWidth,mark repeat={4}, mark options={solid}]
table {%
0.021 0.464421052631579
0.021 0.464421052631579
0.021 0.464421052631579
0.021 0.464421052631579
0.021 0.464421052631579
0.021 0.464421052631579
0.021 0.464421052631579
0.021 0.464421052631579
0.021 0.464421052631579
0.0634 0.620210526315789
0.0634 0.620210526315789
0.1409 0.746421052631579
0.1409 0.746421052631579
0.2635 0.865263157894737
0.3364 0.917894736842105
};
\addplot [thick, green!50.0!black, dashed, mark=triangle*, mark size=\MarkerSize,line width=\LineWidth,mark repeat={4}, mark options={solid,rotate=180}]
table {%
0 0.0186315789473684
5e-05 0.0338947368421053
0.00015 0.0608421052631579
0.0003 0.102421052631579
0.00045 0.167684210526316
0.0007 0.257263157894737
0.00095 0.36778947368421
0.00165 0.483157894736842
0.0027 0.597473684210526
0.00485 0.693157894736842
0.0083 0.773368421052632
0.01355 0.837684210526316
0.0253 0.897578947368421
0.04455 0.938526315789474
0.0633 0.955473684210526
};
\addplot [thick, red, dashed, mark=otimes*, mark size=\MarkerSize,line width=\LineWidth,mark repeat={4}, mark options={solid}]
table {%
0.07765 0.859368421052632
0.07765 0.859368421052632
0.07765 0.859368421052632
0.07765 0.859368421052632
0.07765 0.859368421052632
0.07765 0.859368421052632
0.07765 0.859368421052632
0.07765 0.859368421052632
0.07765 0.859368421052632
0.1197 0.895263157894737
0.1197 0.895263157894737
0.16235 0.922210526315789
0.16235 0.922210526315789
0.2068 0.947789473684211
0.2285 0.961578947368421
};
\addplot [thick, color0, dashed, mark=triangle*, mark size=\MarkerSize,line width=\LineWidth,mark repeat={18}, mark options={solid}]
table {%
0.004 0.937894736842105
0.00425 0.942105263157895
0.00495 0.946315789473684
0.00575 0.950947368421053
0.0066 0.954315789473684
0.0074 0.957368421052632
0.0079 0.959789473684211
0.00935 0.962315789473684
0.01105 0.965894736842105
0.0129 0.968315789473684
0.015 0.970842105263158
0.01785 0.973157894736842
0.0225 0.977157894736842
0.02905 0.980421052631579
0.03405 0.985052631578947
0.03405 0.985052631578947
0.03405 0.985052631578947
0.03405 0.985052631578947
0.03405 0.985052631578947
0.03405 0.985052631578947
0.03405 0.985052631578947
0.03405 0.985052631578947
0.03405 0.985052631578947
0.03405 0.985052631578947
0.03405 0.985052631578947
0.03405 0.985052631578947
0.03405 0.985052631578947
0.03405 0.985052631578947
0.03405 0.985052631578947
0.03405 0.985052631578947
};
\addplot [thick, brown!50.0!black, dashed, mark=halfcircle*, mark size=\MarkerSize,line width=\LineWidth,mark repeat={18}, mark options={solid}]
table {%
0.002 0.665
0.003 0.691
0.004 0.749
0.004 0.721
0.005 0.775
0.007 0.8
0.008 0.82
0.01 0.841
0.013 0.863
0.018 0.882
0.024 0.905
0.034 0.922
0.049 0.944
0.064 0.962
0.068 0.968

};
\addplot [thick, color1, dashed, mark=diamond*, mark size=\MarkerSize,line width=\LineWidth,mark repeat={4}, mark options={solid}]
table {%
0.00795 0.574315789473684
0.00985 0.601157894736842
0.0119 0.629052631578947
0.01495 0.654947368421053
0.01825 0.67978947368421
0.0218 0.701894736842105
0.02745 0.725894736842105
0.03235 0.749473684210526
0.03995 0.772210526315789
0.05005 0.796315789473684
0.06305 0.82221052631579
0.0828 0.855789473684211
0.1158 0.896526315789474
0.1587 0.931789473684211
0.17095 0.941789473684211
};
\addplot [semithick, color2, dashed]
table {%
0 0
0.3364 0.3364
};
\end{axis}

\end{tikzpicture}
          \vspace{-1.0em}
          \caption{Non-defended dataset}          
          \label{fig:ROC_OW_NoDef}
        \end{subfigure}\hspace{0.37cm}
        \begin{subfigure}[ht]{.34\textwidth}
%
%
%
\begin{tikzpicture}

\definecolor{color1}{rgb}{0.75,0.75,0}
\definecolor{color0}{rgb}{0.75,0,0.75}
\definecolor{color2}{rgb}{0,0.75,0.75}

\begin{axis}[
xlabel={False positive rate},
xmin=0, xmax=0.6623,
ymin=0, ymax=1,
width=\smallfigurewidth,
height=180,
xmajorgrids,
x grid style={white!69.01960784313725!black},
ymajorgrids,
legend columns=2,
legend style={at={(0.98,0.29),font=\tiny}},
legend entries={{kNN},{CUMUL},{kFP},{DF},{AWF},{SDAE},{Baseline}},
legend cell align={left}
]
\addplot [thick, blue, dashed, mark=square*, mark size=\MarkerSize,line width=\LineWidth,mark repeat={3}, mark options={solid}]
table {%
0.00125 0.00315789473684211
0.00125 0.00315789473684211
0.00125 0.00315789473684211
0.00125 0.00315789473684211
0.00125 0.00315789473684211
0.00125 0.00315789473684211
0.00125 0.00315789473684211
0.00125 0.00315789473684211
0.00125 0.00315789473684211
0.00795 0.0152631578947368
0.00795 0.0152631578947368
0.0355 0.0514736842105263
0.0355 0.0514736842105263
0.14555 0.178315789473684
0.14555 0.178315789473684
0.14555 0.178315789473684
0.14555 0.178315789473684
0.22 0.25
0.22 0.25
0.22 0.25
0.22 0.25
0.22 0.25
0.6623 0.731789473684211
};
\addplot [thick, green!50.0!black, dashed, mark=triangle*, mark size=\MarkerSize,line width=\LineWidth,mark repeat={4}, mark options={solid,rotate=180}]
table {%
0 0.000421052631578947
0 0.00115789473684211
5e-05 0.00178947368421053
0.0002 0.00378947368421053
0.0003 0.00547368421052631
0.0004 0.0112631578947368
0.00065 0.0196842105263158
0.00095 0.0268421052631579
0.0015 0.0386315789473684
0.00255 0.0558947368421053
0.00615 0.0796842105263158
0.01265 0.123473684210526
0.0294 0.198631578947368
0.09235 0.364526315789474
0.25905 0.627157894736842
};
\addplot [thick, red, dashed, mark=otimes*, mark size=\MarkerSize,line width=\LineWidth,mark repeat={4}, mark options={solid}]
table {%
0.09 0.436105263157895
0.09 0.436105263157895
0.09 0.436105263157895
0.09 0.436105263157895
0.09 0.436105263157895
0.09 0.436105263157895
0.09 0.436105263157895
0.09 0.436105263157895
0.09 0.436105263157895
0.1549 0.550526315789474
0.1549 0.550526315789474
0.24065 0.668526315789474
0.24065 0.668526315789474
0.36085 0.79978947368421
0.48265 0.908526315789474
};
\addplot [thick, color0, dashed, mark=triangle*, mark size=\MarkerSize,line width=\LineWidth,mark repeat={4}, mark options={solid}]
table {%
0.01325 0.683684210526316
0.01605 0.699789473684211
0.01885 0.718736842105263
0.022 0.736
0.02665 0.753263157894737
0.0323 0.770526315789474
0.0389 0.785473684210526
0.04645 0.802
0.05795 0.817157894736842
0.07175 0.834631578947368
0.08915 0.854631578947368
0.114 0.878210526315789
0.1569 0.912421052631579
0.20785 0.946947368421053
0.22885 0.964105263157895
};
\addplot [thick, brown!50.0!black, dashed, mark=halfcircle*, mark size=\MarkerSize,line width=\LineWidth,mark repeat={4}, mark options={solid}]
table {%
0.001	0.059
0.001	0.047
0.001	0.039
0.002	0.072
0.003	0.087
0.005	0.104
0.007	0.123
0.011	0.151
0.016	0.18
0.026	0.221
0.041	0.274
0.068	0.343
0.124	0.447
0.252	0.63
0.398	0.816

};
\addplot [thick, color1, dashed, mark=diamond*, mark size=\MarkerSize,line width=\LineWidth,mark repeat={4}, mark options={solid}]
table {%
0.02135 0.0553684210526316
0.02575 0.0637894736842105
0.0318 0.0736842105263158
0.03945 0.0863157894736842
0.04815 0.0990526315789474
0.05885 0.115368421052632
0.0726 0.135789473684211
0.09055 0.161473684210526
0.11515 0.191684210526316
0.14505 0.234
0.1864 0.285789473684211
0.2475 0.357789473684211
0.3465 0.469789473684211
0.49925 0.646
0.6094 0.770315789473684
};
\addplot [semithick, color2, dashed]
table {%
0 0
0.6623 0.6623
};
\end{axis}

\end{tikzpicture}
          \caption{WTF-PAD dataset}
          \label{fig:ROC_OW_WTF}          
        \end{subfigure}\hspace{-0.2cm}        
        \begin{subfigure}[ht]{.34\textwidth}
%
%
%
\begin{tikzpicture}

\definecolor{color1}{rgb}{0.75,0.75,0}
\definecolor{color0}{rgb}{0.75,0,0.75}
\definecolor{color2}{rgb}{0,0.75,0.75}

\begin{axis}[
xlabel={False positive rate},
xmin=0, xmax=0.82725,
ymin=0, ymax=1,
width=\smallfigurewidth,
height=180,
xmajorgrids,
x grid style={white!69.01960784313725!black},
ymajorgrids,
y grid style={white!69.01960784313725!black},
legend columns=2,
legend entries={{kNN},{CUMUL},{kFP},{DF},{AWF},{SDAE},{Baseline}},
legend style={at={(0.98,0.29),font=\tiny}},
legend cell align={left}
]
\addplot [thick, blue, dashed, mark=square*, mark size=\MarkerSize,line width=\LineWidth,mark repeat={4}, mark options={solid}]
table {%
0.0047 0.00622222222222222
0.0047 0.00622222222222222
0.0047 0.00622222222222222
0.0047 0.00622222222222222
0.0047 0.00622222222222222
0.0047 0.00622222222222222
0.0047 0.00622222222222222
0.0047 0.00622222222222222
0.0047 0.00622222222222222
0.0047 0.00622222222222222
0.0047 0.00622222222222222
0.0047 0.00622222222222222
0.0243 0.0263333333333333
0.0243 0.0263333333333333
0.0243 0.0263333333333333
0.0243 0.0263333333333333
0.0243 0.0263333333333333
0.0821 0.0866666666666667
0.0821 0.0866666666666667
0.0821 0.0866666666666667
0.2221 0.229555555555556
0.2222 0.229555555555556
0.22205 0.229222222222222
0.5486 0.561666666666667
0.55025 0.561666666666667
0.54805 0.556333333333333
0.69815 0.699444444444444
0.69855 0.705888888888889
0.69655 0.700777777777778
0.69695 0.700444444444444
0.6978 0.705777777777778
0.69805 0.700333333333333
};
\addplot [thick, green!50.0!black, dashed, mark=triangle*, mark size=\MarkerSize,line width=\LineWidth,mark repeat={4}, mark options={solid,rotate=180}]
table {%
0 0
0 0
0 0
0 0
0 0
0 0
0 0
0 0
5e-05 0
5e-05 0
5e-05 0
5e-05 0
0.00015 0
0.00025 0.000111111111111111
0.0003 0.000555555555555556
0.0007 0.001
0.0007 0.001
0.0019 0.002
0.0026 0.00288888888888889
0.00455 0.00455555555555556
0.0051 0.005
0.0105 0.0128888888888889
0.0111 0.0137777777777778
0.0262 0.0292222222222222
0.048 0.0538888888888889
0.05905 0.0664444444444444
0.09325 0.0997777777777778
0.0954 0.101777777777778
0.09595 0.102222222222222
0.0964 0.102777777777778
0.09645 0.102777777777778
0.09645 0.102777777777778
};
\addplot [thick, red, dashed, mark=otimes*, mark size=\MarkerSize,line width=\LineWidth,mark repeat={4}, mark options={solid}]
table {%
0.09705 0.104666666666667
0.09705 0.104666666666667
0.09705 0.104666666666667
0.09705 0.104666666666667
0.09705 0.104666666666667
0.09705 0.104666666666667
0.09705 0.104666666666667
0.09705 0.104666666666667
0.09705 0.104666666666667
0.09705 0.104666666666667
0.09705 0.104666666666667
0.09705 0.104666666666667
0.1601 0.172333333333333
0.1601 0.172333333333333
0.1601 0.172333333333333
0.1601 0.172333333333333
0.1601 0.172333333333333
0.2531 0.263333333333333
0.2531 0.263333333333333
0.2531 0.263333333333333
0.4111 0.422333333333333
0.4103 0.422777777777778
0.4107 0.422555555555556
0.65595 0.670444444444444
0.6559 0.671
0.65635 0.67
0.7407 0.750666666666667
0.7394 0.749444444444444
0.742 0.750333333333333
0.742 0.750444444444444
0.7395 0.752777777777778
0.74085 0.752888888888889
};
\addplot [thick, color0, dashed, mark=triangle*, mark size=\MarkerSize,line width=\LineWidth,mark repeat={4}, mark options={solid}]
table {%
0.0293 0.0293333333333333
0.0447 0.0468888888888889
0.06605 0.068
0.09045 0.0975555555555556
0.1214 0.134444444444444
0.1574 0.178444444444444
0.1983 0.228888888888889
0.2411 0.279888888888889
0.28765 0.341555555555556
0.3356 0.405777777777778
0.3875 0.478222222222222
0.45095 0.561333333333333
0.53335 0.661888888888889
0.67015 0.814666666666667
0.77465 0.923444444444444
};
\addplot [thick, brown!50.0!black, dashed, mark=halfcircle*, mark size=\MarkerSize,line width=\LineWidth,mark repeat={4}, mark options={solid}]
table {%
0.002	0.003
0.004	0.005
0.007	0.008
0.011	0.012
0.015	0.019
0.02	0.028
0.029	0.041
0.042	0.057
0.059	0.08
0.083	0.113
0.115	0.153
0.164	0.22
0.248	0.323
0.427	0.52
0.633	0.736

};
\addplot [thick, color1, dashed, mark=diamond*, mark size=\MarkerSize,line width=\LineWidth,mark repeat={4}, mark options={solid}]
table {%
0.0033 0.00322222222222222
0.00345 0.00333333333333333
0.0037 0.00366666666666667
0.0041 0.00411111111111111
0.0044 0.00433333333333333
0.005 0.00466666666666667
0.00535 0.00466666666666667
0.00585 0.00522222222222222
0.0063 0.006
0.00665 0.00655555555555556
0.0077 0.00755555555555556
0.00875 0.00877777777777778
0.0108 0.0111111111111111
0.01485 0.0156666666666667
0.0376 0.0463333333333333
};
\addplot [semithick, color2, dashed]
table {%
0 0
0.82725 0.82725
};
\end{axis}

\end{tikzpicture}
          \caption{W-T dataset}
          \label{fig:ROC_OW_WT}
        \end{subfigure} 
        \vspace{-0.3em}
         \caption{ROC curve in Open-world scenario}\label{fig:open-world-roc}
        \vspace{-0.9em}
\end{figure*}

We plot the ROC curve for all the WF attacks against non-defended, WTF-PAD and W-T datasets using the standard model in the open-world scenario as shown in Figures~\ref{fig:ROC_OW_NoDef}$-$\ref{fig:ROC_OW_WT}. The ROC curve allows us to evaluate the classifier and strive for a trade-off between TPR and FPR. For example, the best overall results for DF against non-defended traffic might be optimizing for high TPR, with 0.98 TPR and 0.03 FPR, and optimizing for low FPR, with 0.94 TPR and 0.004 FPR. 

\end{document}